\documentclass[amsmath,amssymb,aps,twocolumn,groupedaddress]{revtex4-1}

\usepackage[breaklinks=true]{hyperref}
\usepackage{graphicx}
\usepackage{braket}
\usepackage{algorithm}
\usepackage{algpseudocode}
\usepackage{xcolor}
\begin{document}
\preprint{APS/123-QED}

\title{Coarse-grained spectral projection (CGSP): a deep learning-assisted approach to quantum unitary dynamics}

\author{Pinchen Xie}
\affiliation {Program in Applied and Computational Mathematics, \\ Princeton University,
NJ 08544, USA}

\author{Weinan E}
\affiliation {Department of Mathematics and Program in Applied and Computational Mathematics, \\ Princeton University,
NJ 08544, USA}

\date{\today}

\begin{abstract}
 We propose the coarse-grained spectral projection method (CGSP), a deep learning-assisted approach for tackling quantum unitary dynamic problems with an emphasis on quench dynamics. We show CGSP can extract spectral components of many-body quantum states systematically with sophisticated neural network quantum ansatz. CGSP exploits fully the linear unitary nature of the quantum dynamics, and is potentially superior to other quantum Monte Carlo methods for ergodic dynamics. Preliminary numerical results on 1D XXZ models with periodic boundary condition are carried out to demonstrate the practicality of CGSP.
\end{abstract}

\maketitle

{ \it Introduction.---}
The past several decades have witnessed a rapid growth of research interests in dynamical quantum many-body systems~\cite{Kaufman2012, Phillips1998, Weiss2017, Spring2013,Vandersypen2004}, leading to the observation of novel quantum phenomena ~\cite{Schmitz2009, Zhang2017a, Zhang2017, Smith2016, Gring2012} outside the scope of equilibrium physics. It also opened the possibility for realistic implementations of quantum computations~\cite{Cai2013,Houck2012,  Harty2014, Arute2019, Pan2012}. 

At the same time, there have also been a great deal of activity on scalable algorithms for numerical simulations of quantum dynamics~\cite{Wahl, Devakul2015,Schroder2019,Khasseh2020,DelPino2018,Werner2016,Doria2011,Bukov2018,Niu2019,Wang2019,Worth2008} . The main challenge in this pursuit is modeling  highly-entangled high-dimensional quantum states present during evolution, a task usually requires exponential complexity in classical computing. 
 Examples that fall in this category include
most tensor network ansatz (including Matrix product states (MPS)), projected entangled pair states (PEPS) and multiscale entanglement renormalization ansatz (MERA)~\cite{Schollwock2011,Verstraete2008,Vidal2007}, originated from the density matrix renormalization group (DMRG) method~\cite{White1992}. As a consequence, the application of these ansatzs is usually limited to 1D/2D systems featuring area-law entanglement with or without logarithmic correction~\cite{Hastings2007, Bravyi2006}. Hence more versatile ansatzs are desired in the face of quantum dynamics.

In recent years, the most promising candidate turned out to be artificial neural networks which are believed to have huge entanglement capacity~\cite{Deng2017}. An early practice along this line of research was the application of restricted Boltzmann machine (RBM) in solving the ground state and the dynamics of quantum spin models~\cite{Carleo2017}. Later, symmetry preserving deep fully-connected neural networks (FNN) and convolutional neural networks (CNN) were also shown to be efficient quantum state ansatz~\cite{han2019solving, Choo2019, Pfau2019,luo2019backflow, hermann2019deep}. In particular,  CNN is believed to support volume-law entanglement scaling while being polynomially more efficient in resources compared to RBM-like ansatz in 2D, due to an inherent reuse of information~\cite{Levine2019}. 

So far, the main algorithm accompanying black-box-like neural networks for simulating quantum dynamics is the time dependent variational principle (TDVP) method~\cite{Carleo2017}. In plain words, TDVP projects real-time quantum evolution trajectory into a tiny useful subset, parameterized by a neural network, of the Hilbert space. The projected dynamics is then described by a low-dimensional time-dependent differential equation of neural network parameters. In spite of the numerical instability and the limited expressive power of the neural networks, TDVP methods are potentially capable of simulating quench dynamics of very large quantum spin systems with strong entanglement~\cite{Schmitt}, ultrafast dynamics~\cite{Fabiani}, the evolution of open quantum systems as well as stationary states~\cite{Hartmann2019,yoshioka2019constructing}.
However, TDVP methods do not give special treatment to dynamics driven by a static Hamiltonian where the quantum evolution has a certain spectral structure and multiple intrinsic time scales. The ignorance of both may lead to prohibitive numerical instability in integrating the TDVP-induced low-dimensional dynamics step-by-step.  

In this work, we will show how to poke into the spectral structure of unitary dynamics and extract limited but useful high-dimensional information directly from initial condition. 
This is done through a coarse-grained representation of the spectral projection with deep learning, a procedure we dubbed  coarse-grained spectral projection (CGSP). The results of CGSP can be used to simulate a unitary dynamics driven by a static Hamiltonian directly without step-by-step integration.


{\it Coarse-grained spectral projection.---}
Considering a pure quantum state $\ket{\Psi_o}$ (in the following the brackets for a ket will be dropped in the absence of inner product) in a closed system as the initial condition of an unitary evolution driven by the static Hamiltonian $H$, a complete eigen-decomposition of $\Psi_o$ can be expressed as 
\begin{equation}\label{basis}
    \Psi_o =\sum_{i=1}^{N_h} b_i \psi_i
\end{equation}
where $\{b_i\}_{i\in[1,N_h]}$ are real constants.
The eigenstates $\{\psi_i\}_{i\in[1,N_h]}$ satisfy $H\psi_i = E_i\psi_i$. They are orthonormal and increasingly ordered with respect to the energy level $E_i$. $N_h$ is the dimension of the Hilbert space $\mathcal{H}$.
There exists trivial disjoint cover of the entire energy spectrum on the real axis:
\begin{equation}\label{split}
[E_{1},E_{N_h}] \subset [x_{0},x_1) \cup [x_1,x_2) \cup \cdots \cup[x_{N-1},x_{N}]    
\end{equation}
such that $\{x_i\}_{i\in[1,N]}$ is an arithmetic sequence satisfying  $x_{0}<E_1<E_{N_h}<x_{N}$. Then for  each interval $[x_i,x_{i+1}]$ we associate the direct sum of the eigen-subspaces whose  eigenvalues lie in $[x_i,x_{i+1}]$. We obtain
 $N$ subspace $\{Q_i\in \mathcal{H}\}_{i\in[0,N-1]} $ and $N$ corresponding projection operators $\{\mathcal{P}_i\}_{i\in[0,N-1]}$.
Since $\mathcal{H} = \mathop{\oplus}\limits_i Q_i$, it is obvious that 
$\Psi_o \in \text{span}(\mathcal{P}_0\Psi_o, \cdots, \mathcal{P}_{N-1}\Psi_o)$.
Let $\theta_i$ denote the normalized $\mathcal{P}_i\Psi_o$. $\Psi_o$ can be expressed as $    \Psi_o = \sum_{i=0}^{N-1} c_i \theta_i$ where $c_i$ are real constants.

Let $\epsilon=x_{i+1}-x_{i}$  and $\lambda_i=(x_{i}+x_{i-1})/2$ be the center of the $i$-th interval. The unitary evolution of $\Psi_o(t)=e^{-iHt}\Psi_o$ driven by time-independent Hamiltonian $H$ can be approximated by 
\begin{equation}\label{dynamics}
    \varphi_o(t) =   \sum_{i=0}^{N-1} c_ie^{-i\lambda_i t} \theta_i.
\end{equation}
The error has an evident upper bound given as $\| \Psi_o(t) - \varphi_o(t) \| \leq \frac{\epsilon t}{2}$. To achieve $\| \Psi_o(t) - \varphi_o(t) \|<\delta$ for small enough $\delta$, $N$ should satisfy 
\begin{equation}\label{error}
    N > \frac{(E_{N_h}-E_1)t}{2\delta}.
\end{equation}
Notice that the energy spectrum range $(E_{N_h}-E_1)$ usually grows linearly with system size for quantum models defined on (nearly) regular graphs with bounded short-range interaction and disorder. 
Recently, a similar bound has also been derived in the context of quantum state compression via principle component analysis~\cite{Kosut2020}.

 It is both unnecessary and difficult, if not impossible, to solve the $N$ projected state $\theta_i$ exactly for simulating dynamics. Compromises should be made for practical reason. Especially, the uniqueness of $\theta_i$ should be loosened by allowing polluted projection, which implies a non-orthogonal decomposition of $\Psi_o(0)$.  Assuming $N$ normalized states $\Theta_i$ parameterized by $N$ classical ansatzs respectively, a non-orthogonal decomposition of $\Psi_o(0)$ can be achieved through the following objective function as a constrained optimization problem:
\begin{equation}\label{obj}
\begin{split}
    \mathop{\text{minimize}}\limits_{\{c_i\},\{\Theta_i\}} \ \ &\sum_{i=0}^{N-1}c_i^2\braket{\Theta_i\vert (H-\Lambda_i)^2\vert\Theta_i},  \\
    \text{subject to } \ \ &d(\Psi_o(0),\sum c_i\Theta_i) = 0. \\
\end{split}
\end{equation}
$d$ can be any legal distance function in the Hilbert space including the Fubini-Study metric and $L_2$ norm, regardless of $U(1)$ symmetry.   
The fixed constants $\{\Lambda_i\}$ in the objective function is an arithmetic sequence satisfying $\Lambda_0 \leq E_{min}$ and $\Lambda_{N-1} \geq E_{max}$, with a common increment of $\epsilon>0$. The ground state energy $E_{min}$ of the system Hamiltonian and the maximum energy $E_{max}$ can be easily estimated by usual VMC techniques with the same classical ansatz.

With a slight abuse of notation, let $\lambda_i = \frac{\braket{\Theta_i\vert H\vert \Theta_i}}{\braket{\Theta_i\vert \Theta_i}}$. $\Psi_o(t)$ can be approximated by $\varphi_o(t)=\sum_{i=0}^{N-1} c_ie^{-i\lambda_i t} \Theta_i$ up to a constant phase difference $\phi_{ph}$. Without loss of generality, we assume $\phi_{ph}=0$ for all that follows.
The quality of this approximation is reflected by the 
``covariance'' matrix
\begin{equation}
    (K)_{ij} = {\braket{\Theta_i\vert (H-\lambda_i)(H-\lambda_j)\vert \Theta_j}}.
\end{equation}
For small $t$, the error of the approximation is 
\begin{equation}\label{aerror}
    \| \Psi_o(t) - \varphi_o(t) \|^2 \approx t^2 {\bf c}^\dagger K{\bf c}.
\end{equation}

In practice, it is easier to only calculate the diagonal element of $K$, i.e. the variance $\sigma_i^2 = \braket{\Theta_i\vert (H-\lambda_i)^2\vert \Theta_i}$ of each $\Theta_i$. 
Let $|\sigma|^2 = \sum_i c_i^2\sigma_i^2/\sum_i c_i^2$.
We have a very rough but inexpensive estimation of the error:
\begin{equation}\label{err}
    \| \Psi_o(t) - \varphi_o(t) \|^2 \approx \vert\sigma\vert^2 t^2.
\end{equation}

Notably, Eq.~(\ref{obj}) minimizes a weighted sum of individual variances. We show in the supplementary material that this particular choice of objective function leads to a $N^{-1}$ scaling of each $\sigma_i$ in the ideal case.

When $N \rightarrow \infty$, Eq.~(\ref{obj}) converges to its continuous form:
\begin{equation}\label{con_obj}
\begin{split}
    \mathop{\text{minimize}}\limits_{c,\Theta} \ \ &\int_{w_a}^{w_b}c^2(w)\braket{\Theta(w)\vert (H-w)^2\vert\Theta(w)}dw,  \\
    \text{subject to } \ \ &d(\Psi_o(0),\int_{w_a}^{w_b}c(w)\Theta(w)dw) = 0. \\
\end{split}
\end{equation}
For many disordered system, eigenstates with very close energy levels can have completely different local observables~\cite{Alet2018}. Hence a global minimizer $\Theta_m(w)$ of Eq.~(\ref{con_obj}) is not expected to be continuous with respect to $w$. Therefore we adopt the discrete form Eq.~(\ref{obj}) as the starting point for extracting spectral information and name it ``coarse-grained spectral projection''.

{\it Numerical framework and results.---}
To show CGSP is applicable to real quantum dynamic problems, we will propose a feasible numerical framework for using Eq.~(\ref{obj}) in quench dynamics of quantum lattice models. When deep neural networks serve as ansatzs, it is desirable to convert Eq.~(\ref{obj}) into an unconstrained loss function for practical training. A naive treatment is to handle the constraint in Eq.~(\ref{obj}) with penalty method. We found this approach very problematic because a noisy estimation of the emphasized penalty will greatly slow down the minimization of the original objective function. A more considerate approach is to construct the ansatzs in a way that the constraint is automatically satisfied. 

Suppose the initial state $\Psi_o(0)\in\mathcal{H}$ can be parameterized exactly by classical ansatz $\Upsilon_0$ with fixed parameters. In addition, we have $M$ classical ansatzs $\{\Upsilon_j\}_{j\in[1,M]}\subset \mathcal{H}$ with free parameters.  Let $A=(A_{ij})_{i\in[0,N-1],j\in[0,M]}$ be a real matrix. 
Then $c_i\Theta_i$ as a whole is constructed to satisfy $d(\Psi_o(0),\sum c_i\Theta_i) = 0$, given as
\begin{equation}\label{theta_cons}
    c_i\Theta_i = \sum_{j=0}^M(\frac{\delta_{j0}}{N}+A_{ij}-\frac{\sum_{i=0}^{N-1}A_{ij}}{N})\Upsilon_j.
\end{equation}
In principle $M$ should be larger than $N$ to ensure the linear independence of $\{\Theta_i\}_{i\in[0,N-1]}$. Also, larger $M$ would provide stronger variational freedom for $\{\Theta_i\}_{i\in[0,N-1]}$.
But in practice $M$ is flexible because some $c_i$ vanishes. 
Next, we  define a new objective function  without explicit constraint:
\begin{equation}\label{tloss}
    L = \Big(\frac{2(N-1)}{\Lambda_{N-1}-\Lambda_0}\Big)^2 \sum_{i=0}^{N-1}c_i^2\braket{\Theta_i\vert (H-\Lambda_i)^2\vert\Theta_i}.
\end{equation}
Eq.~(\ref{tloss}) is used for training  $\{\Upsilon_j\}_{j\in[1,M]}$ and $A$. The constant in front of the summation in Eq.~(\ref{tloss}) ensures that the minimum of $L$ is in the order of $O(1)$ rather than $O(N^{-2})$. 

Due to exponentially large Hilbert space, $L$ should be estimated with Monte Carlo methods. Compared to traditional sequential Monte Carlo sampling methods such as Markov Chain Monte Carlo, we find recently developed neural autoregressive quantum states (NAQS) ~\cite{Sharir2020} can achieve higher efficiency and better sampling quality at the same time, if employed on graphics processing units(GPUs). Moreover, NAQS allows exact normalization. So in the following numerical results, we use our CGSP-adapted NAQS as classical ansatzs and the direct sampling algorithm associated to NAQS for Monte Carlo sampling. This CGSP-adapted NAQS supports parallel evaluation of $\{\Upsilon_j\}_{j\in[1,M]}$ and also parallel sampling. Detailed information can be found in the supplementary material. 
\begin{figure}[t]
    \centering
    \includegraphics[width=\linewidth]{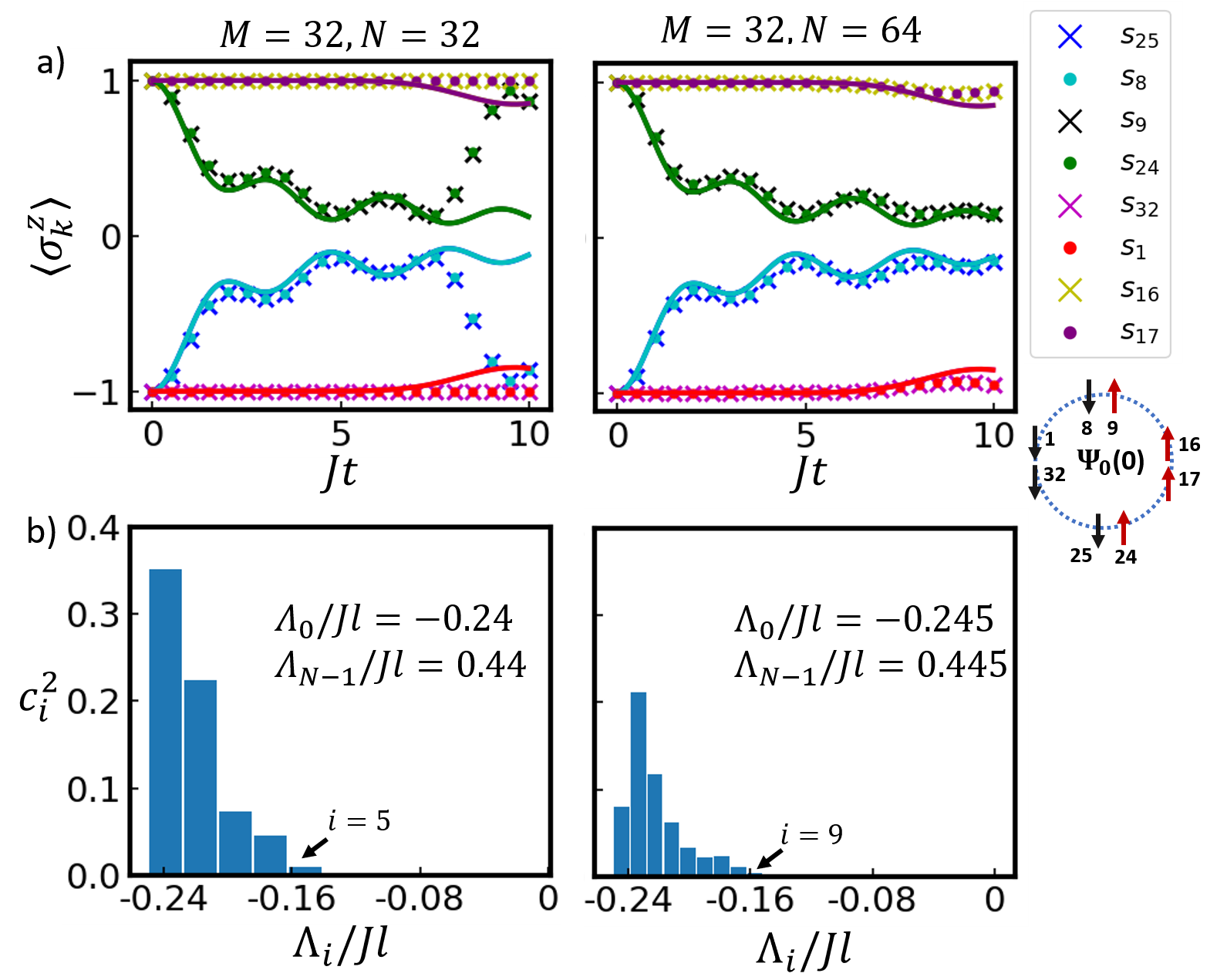}
    \caption{(a) Dynamics of local magnetization  $\braket{\sigma^z_k(t)}$ for spins adjacent to the domain wall and spins at the middle of the ferromagnetic domain (the cartoon below the legend specifies the labeling). The scattered data are calculated by CGSP with $(M,N)=(32,32)$ (left column) and $(M,N)=(32,64)$ (right column).
    The solid lines are results from converged TDVP-MPS. The color of the solid line matches the color of scattered data for the same spin. Due to the mirror symmetry of the initial state, the solid lines associated to $k=9,16,25,32$ are covered by the others.
    (2) The amplitude $c_i^2$ of projected states for $(M,N)=(32,32)$ (left column) and $(M,N)=(32,64)$ (right column). Because $c_i^2$ vanishes for all $\Lambda_i>0$, we plot only the lower section of the energy spectrum. }
    \label{fig:cgsp-cgsp}
\end{figure}
\begin{figure}[t]
    \centering
    \includegraphics[width=\linewidth]{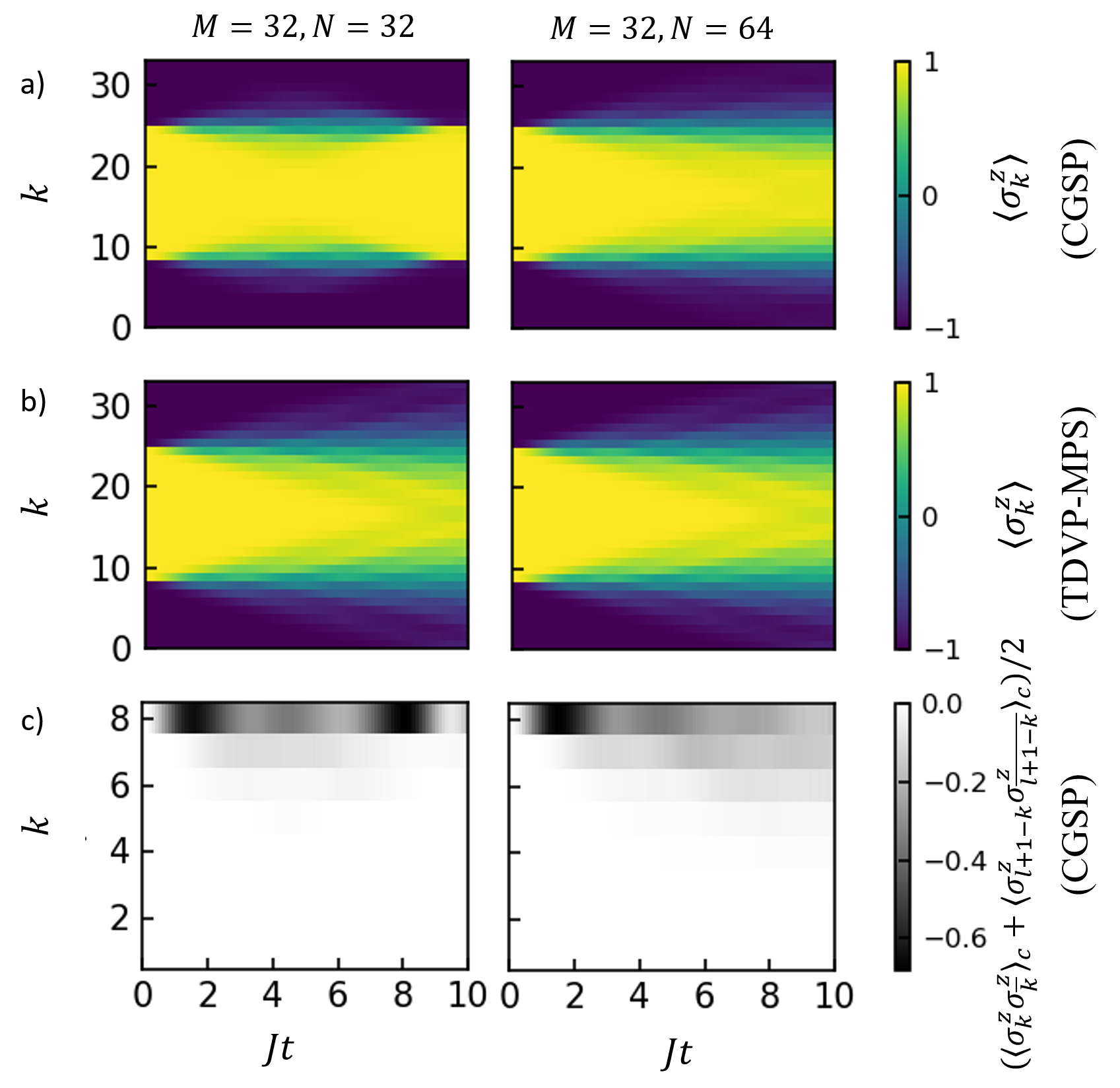}
    \caption{(a) Dynamics of local magnetization $\braket{\sigma^z_k(t)}$ after a sudden quench calculated by CGSP for the 32-spin XXZ model. (b) Exact dynamics of $\braket{\sigma^z_k(t)}$ obtained with TDVP-MPS. The right column is a duplicate of the left column for easier comparison. (c) The correlation function $\braket{\sigma_k^z\sigma_{\overline{k}}^z}_c$ between spin $k$ and its counterpart spin $\overline{k}$ ($\overline{k}=\frac{l}{2}+1-k$), averaged with $\braket{\sigma_{l+1-k}^z\sigma_{\overline{l+1-k}}^z}_c$.  }
    \label{fig:cgsp-tdvp}
\end{figure}

In the following, we will demonstrate the practicality of CGSP by simulating the unitary quench dynamics of 1D spin-$1/2$ XXZ model. The Hamiltonian is given by
\begin{equation}
    H(J,\Delta,h)=\sum_{k=1}^l J(S_k^xS_{k+1}^x + S_k^yS_{k+1}^y + \Delta S_k^zS_{k+1}^z) + hS_k^z.
\end{equation}
We assume periodic boundary condition and work strictly within the zero total $S^z$ sector so $h$ is irrelevant. 
For the numerical results presented here, the XXZ chain contains $l=32$ spins suddenly quenched from $\Delta\rightarrow-\infty$ to $\Delta=-1$ with initial condition $\Psi_0(0)=\ket{\downarrow\downarrow\cdots\downarrow\uparrow\uparrow\cdots\uparrow}$. We compare our results to converged TDVP calculation with MPS (TDVP-MPS). In Fig.~\ref{fig:cgsp-cgsp}(a), we plot the $\sigma^z$ local magnetization of several representative spins computed by CGSP with $(M,N)=(32,32)$ and $(M,N)=(32,64)$ respectively. In Fig.~\ref{fig:cgsp-cgsp}(b), we show the amplitude of non-vanishing projected states. It is evident from Fig.~\ref{fig:cgsp-cgsp}(b) that the initial state $\Psi_0(0)$ contains mainly low-lying eigenmodes of $H(J,-1, h)$. This  justifies using CGSP with $M<N$ for $\Psi_0(0)$. Based on this observation, $M=32$ should be 
 enough for CGSP with $N=32$ and $N=64$. In Fig.~\ref{fig:cgsp-cgsp}(a), we find with $N=64$ CGSP can simulate longer dynamics than $N=32$. If we increase $N$ further, CGSP should be more accurate until the expressive power limited by $M=32$ becomes the main bottleneck. 

In Fig.~\ref{fig:cgsp-tdvp}, we show the dynamics 
of $\braket{\sigma^z_k(t)}$ for all the spins (Fig.~\ref{fig:cgsp-tdvp}(a)), compared to the TDVP-MPS benchmarks (Fig.~\ref{fig:cgsp-tdvp}(b)). For CGSP simulation with $(M,N)=(32,64)$, the evolution of the local magnetization shows the light-cone structure predicted by the Lieb-Robinson bounds. In Fig.~\ref{fig:cgsp-tdvp}(c), we plot the correlation function, calculated by CGSP, between pairs of spins with the same distance from the domain wall but in opposite sides according to the initial state. Long-range correlation emerges during evolution.

Based on Eq.~(\ref{err}), the numerical coherence time $T_c$ with respect to $\| \Psi_o(T_c) - \varphi_o(T_c) \|^2 \approx 0.5$ can be estimated from the training results for predicting the valid region of simulated dynamics without benchmarking. We obtain $JT_c \approx 3.5$ for $(M,N)=(32,32)$ and $JT_c\approx 5.6$ for $(M,N)=(32,64)$. By observing  Fig.~\ref{fig:cgsp-cgsp} and Fig.~\ref{fig:cgsp-tdvp}, we see that $T_c$ may slightly underestimate the region of validity of the simulated dynamics.

Technical details of numerical experiments can be found in the supplementary material, where we also propose a simple parallel framework for breaking down CGSP into hierarchically-organized sub-tasks.

{\it Discussion.---}
Our numerical experiments mainly showcase the practicality of CGSP without further analyzing its scalability and other issues such as entanglement, symmetry, non-locality, thermalization, etc. Neither do we show how different types of classical ansatz can be fitted into the framework of CGSP. But this does not prevent us from estimating the complexity of CGSP in terms of ansatz complexity and quantum system specifications. Suppose the spectrum range of a $l$-spin Hamiltonian with only short-range interaction is $E$ and the time scale we want to simulate is represented by $T$. The number of necessary  projected states  obeys $O(ET)$. The number of stochastic samples needed to control noise level is $O(T^4)$. So the computational cost for estimating the loss function (including the sampling process) is $O(lT^4)\times C(l,ET)$ where $C(l,ET)$ denotes the computational complexity of one forward propagation of the classical ansatz in terms of $l$ and $ET$. For optimization methods based on first-order gradient descent, the total number of iterations required for convergence is unknown. 

Unfortunately, even though we guess $C(l,ET)$ to be polynomial for some specific tasks, their is no conclusive complexity theory yet to predict $C(l,ET)$ or the neural network complexity in other numerical algorithms. This renders the comparison between deep learning algorithms rather difficult, especially when there is no general-purpose neural network structure for different kinds of quantum problems. But we are still able to make some qualitative comments. It is helpful to recall that tensor network ansatzs have almost sure polynomial complexity in some fully many-body localized systems~\cite{wahl2017efficient}, due to its faithful representation of weakly-entangled regime inside the Hilbert space. 
However, when we consider TDVP-based evolution of neural networks for generic systems, it is not harbored by weakly-entangled regime. Though carrying high entanglement capacity, a finite-size classical ansatz can only represent a low-dimensional section of the Hilbert space. For TDVP methods, polynomial complexity is not possible when the actual quantum state trajectory travels away from the low-dimensional section. This failure is inevitable for ergodic dynamics harnessed by few symmetries and may be more easily detected in quench dynamics over criticality~\cite{Czischek2018}.
For CGSP, the breakdown of polynomial complexity is another scenario. Notably, in CGSP, the neural network is expected to parameterize only $O(T)$ quantum many-body states rather than a differentiable subset containing the real-time evolution trajectory. This statement holds true, regardless of ergodicity, for finite-time dynamics driven by time-independent Hamiltonian. Nevertheless, when $T$ is large or the target quantum state is featureless, even a countable finite subset of Hilbert space is too difficult for neural networks to represent fully. This is when CGSP also encounters exponential complexity. 

Based on the discussion above, it seems that CGSP has a lighter burden to bear for quench dynamics, with possibly only a $O(t)$ scaling factor on top of the complexity of ground states represented in the ansatz. However, from the optimization perspective, CGSP requires more training efforts compared to TDVP methods which propagate in a deterministic way when Monte Carlo sampling is nearly exact. Because the optimization of a CGSP task is non-convex towards the objective Eq.~(\ref{obj}), CGSP may suffer from local optimum and ill-conditioning like almost every deep learning task. Since gradient-based optimization methods are almost the only practical choice for deep neural network, these issues can be the major obstruction against the scalability of CGSP.

{\it Outlook.---}
So far, we find CGSP to be potentially a good candidate for studying the unitary dynamics of quantum systems, for it not only provides the access to almost all observables but also unfolds the spectral structure of an unitary evolution. More meaningful physics are encoded in the CGSP results than conventional VMC simulations. Being fundamentally different from previous methods utilizing TDVP or Krylov subspace, CGSP is expected to solve specific problems that are inaccessible in the past. There is also the possibility that CGSP can improve TDVP simulations driven by slow-varying time-dependent Hamiltonian. The details of the latter is included in the supplementary material.  

Future development of CGSP may focus on more efficient utilization of the spectral structure of an initial state, or designing more sophisticated loss function for enhancing the orthogonality between projected states. It is also possible to apply CGSP to the unitary dynamics of molecular systems. Moreover, a lot of efforts should be devoted to further developing neural network ansatz that can model quantum states in different scenarios, for example states near thermalization. 

{\it Acknowledgement.---} This work is supported in part by a gift to Princeton University from
iFlytek.

{\it Code Availability.---} The codes for the implementation of CGSP-adapted NAQS and the numerical experiments are available at
\href{https://github.com/salinelake/cgsp}{\nolinkurl{https://github.com/salinelake/cgsp}}.

\appendix
\section{Ideal minimizer of Objective function}\label{ax1}
We are going to derive the global minimizer of Eq.~(\ref{obj}) of the main text in the ideal case that the classical ansatz $\{\Theta_i\}_{i\in[0,N-1]}$ can represent any quantum states faithfully. We will use the same notation $\{\psi_i\}_{i\in[1,N_h]}$ to denote an increasingly ordered orthonormal eigen-basis associated to Hamiltonian $H$ as in Eq.~(\ref{basis}) of the main text.

For each $\Theta_i$, its unique eigen-decomposition can be expressed as
\begin{equation}
    \Theta_i = \sum_{k=1}^{N_h} a^{(i)}_k\psi_k
\end{equation}
In the same way, the initial condition can be decomposed into
\begin{equation}
    \Psi_o(0) = \sum_{k=1}^{N_h} b_k\psi_k.
\end{equation}

Using $L_2$ norm as the distance measure, the original optimization problem (Eq.~(\ref{obj}) of the main text) can be written as 
\begin{equation}\label{aobj}
\begin{split}
    \mathop{\text{minimize }}\limits_{\{c_i\},\{a_k^{(i)}\}} \ \ &\sum_{i=0}^{N-1} \sum_{k=1}^{N_h} \vert c_ia^{(i)}_k\vert^2(E_k-\Lambda_i)^2,  \\
    \text{subject to } \ \ & \sum_{k=1}^{N_h}\vert b_k - \sum_{i=0}^{N-1} c_ia^{(i)}_k \vert^2=0, \\
\end{split}
\end{equation}
Let $g_{i,k} = c_ia^{(i)}_k$, the necessary conditions for the minimizer $(\overline{g}_{i,k} = \overline{c}_i\overline{a}^{(i)}_k)$ can be written as
\begin{equation}\label{lag1}
    \overline{g}_{i,k}(E_k-\Lambda_i)^2 = \mu_k
\end{equation}
and
\begin{equation}\label{lag2}
    b_k-\sum_{i=0}^{N-1} \overline{g}_{i,k}=0.
\end{equation}
$\mu_k$ is an undetermined multiplier in Eq.~(\ref{lag1}).
Combining Eq.~(\ref{lag1}) and Eq.~(\ref{lag2}) yields
\begin{equation}\label{cct}
    \overline{g}_{i,k} = \frac{b_i}{\sum_{j=0}^{N-1}\frac{(E_k -\Lambda_i)^2}{(E_k -\Lambda_j)^2}}
\end{equation}
To understand Eq.~(\ref{cct}), recall the definition in the main text that $\{\Lambda_i\}$ is evenly spaced with energy gap $\epsilon$. We call $\overline{d}_{i,k}=\vert E_k -\Lambda_i\vert/\epsilon$ the relative spectral distance between the $k$-th eigenmode and $\overline{\Theta}_i=\sum \overline{a}_k^{(i)}\psi_k$. In addition, we interpret $\vert \overline{g}_{i,k}/b_i\vert^2$ as the dispersion of the $k$-th eigenmode in the minimizer.
When $(\Lambda_{N-1} - \Lambda_0)$ is fixed and $N$ is large enough, Eq.~(\ref{cct}) suggests  $\vert \overline{g}_{i,k}/b_i\vert^2$ scale with $1/\overline{d}_{i,k}^{2}$, which also leads to that
$\frac{\braket{\overline{\Theta}_i\vert (H-\Lambda_i)^2\vert\overline{\Theta}_i}}{\braket{\overline{\Theta}_i\vert\overline{\Theta}_i}}$ scales with $\epsilon^2$. Hence we can conclude that the particular choice of objective function decided by Eq.~(\ref{obj}) of the main text can systematically improve the monochromaticity of each $\Theta_i$ by squeezing the dispersion of every eigenmode $\psi_k$.

\section{CGSP-adapted neural autoregressive quantum states}
In addition to serving as eligible VMC ansatzs,  neural autoregressive quantum states (NAQS) can greatly boost the efficiency of stochastic importance sampling. Before NAQS, the sampling tool accompanying neural quantum states is Markov-Chain Monte-Carlo (MCMC) by default. Though parallelizable to some extent, MCMC is essentially a sequential algorithm. The fact MCMC requiring a long mixing time is not prefered by large-scale deep learning applications using graphics processing units (GPUs). In contrast, NAQS realize importance sampling in a parallel manner suitable for GPUs. 
\begin{figure}[tb]
    \centering
    \includegraphics[width=0.8\linewidth]{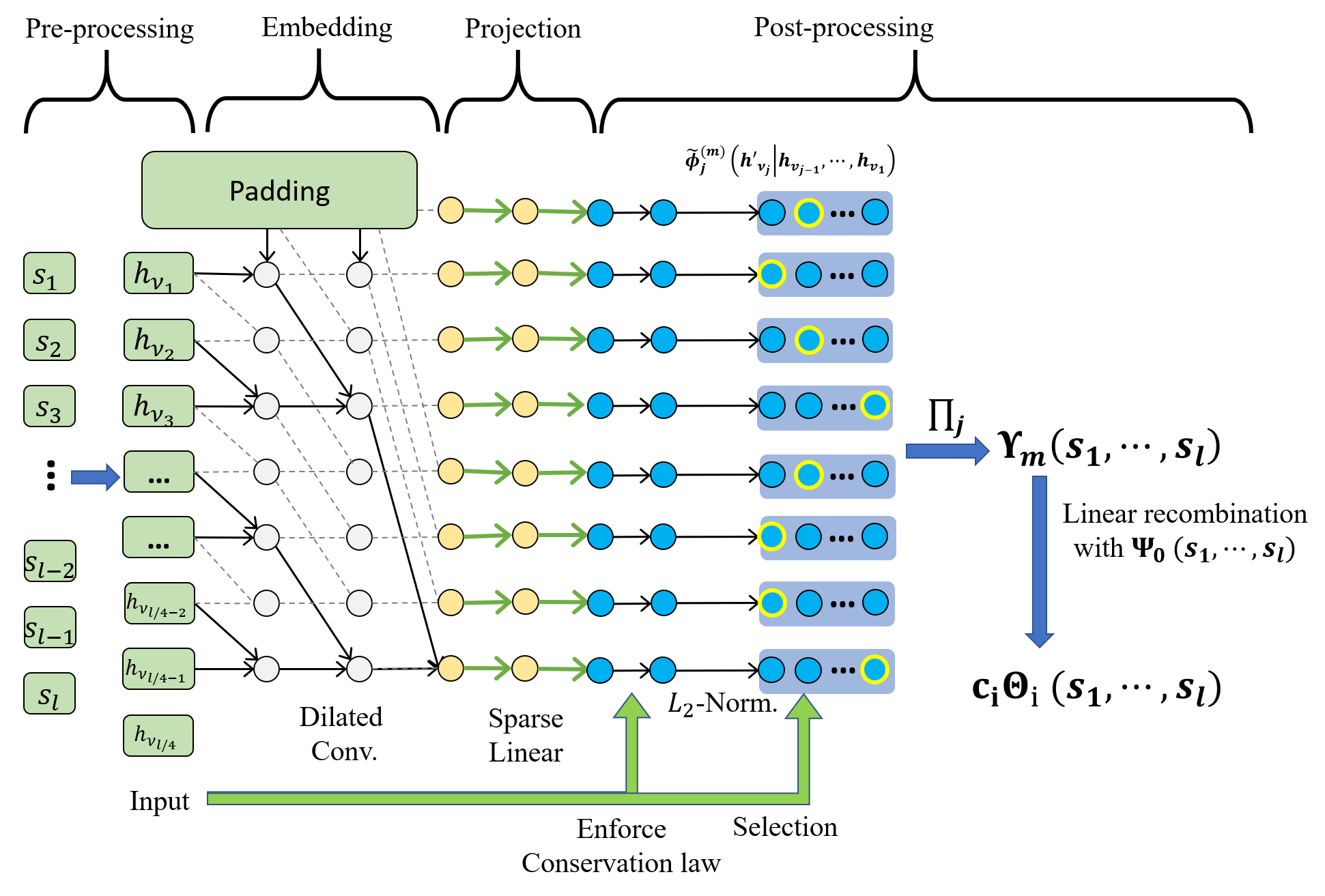}
    \caption{Structure of a CGSP-adapted NAQS and the workflow of forward propagation in the evaluation mode. The original input is a batch of 1D spin configurations. The output is a batch of complex $N$-vectors, i.e. $\{\Theta_k(s_1,\cdots,s_l)\}_{k\in[0,N-1]}$.  The forward propagation consists of 4 steps. (i) Pre-processing: Reordering the original spin configuration and regrouping the outcome into its hexadecimal equivalent. (ii) Embedding: Dilated convolution layers. Only one end of the convolution input is padded with zero such that the convolution output obeys the conditional dependence required by NAQS. (iii) Projection: Fully connected layers applied to each input node respectively. The connection is sparse for the entire input tensor . Let $N_B$ denote the batch size. The output is a 4D (5D) real tensor of size ($M$, $N_B$, $l/4$, 16, (2)). The fourth dimension is associated to $[0,1]^4$, i.e. the possible outcomes of $h'_{\nu_j}$. The optional 5-th dimension is devoted to representing the real part and the imaginary part of a complex wave function separately. (iv) Post-processing. Conservation law is enforced by multiplying tensor element corresponding to illegal spin configurations with zero. $L_2$-normalization $x\rightarrow x/\sqrt{||x||_2}$ is imposed for evaluating the conditional wave function $\tilde{\phi}_j^{(m)}$ at possible configurations. Then with the input spin configuration, one can select the corresponding outcomes from the conditional wave function and compute the total wave function $\Upsilon_m(s_1,\cdots,s_l)$ ($m\in[1,M]$). Finally, with Eq.~(\ref{theta_cons}) of the main text, $\{\Upsilon_m(s_1,\cdots,s_l)\}_{m\in[1,M]}$ together with the initial condition are recombined to produce the final results.}
    \label{fig-naqs}
\end{figure}
A NAQS is a normalized wave function that can be expressed as a product of normalized conditional wave function. For a general introduction to NAQS readers are refered to Ref.~\cite{Sharir2020}. Here we will only give the example of NAQS in the context of spin-$1/2$ models. With the $S_z$ basis of a 1D XXZ model, a NAQS can be expressed as  
\begin{equation}\label{naqs_eq}
    \Upsilon(s_1,\cdots,s_l) = \prod_{i=1}^{l} \phi_i(s_{\mu_i}\vert s_{\mu_{i-1}},\cdots,s_{\mu_1}) 
\end{equation}
where $(\mu_1,\cdots,\mu_l)$ is a permutation of the natural spin order $(1,\cdots, l)$ in a 1D chain. The conditional wave function $\phi_i$ should satisfy a local normalization condition
\begin{equation}\label{cond-norm}
    \sum_{s'_{\mu_i}\in\{\downarrow,\uparrow\}} \|\phi_i(s'_{\mu_i}\vert s_{\mu_{i-1}},\cdots,s_{\mu_1})\|^2
    =1
\end{equation} 
for any legal configuration $(s_1,\cdots,s_l)$ that does not break any conservation law. 
With Eq.~(\ref{naqs_eq}) and Eq.~(\ref{cond-norm}), the wave function $\Upsilon$ is automatically normalized. When $\Upsilon$ is inside a specific $S_z$ sector, any $\phi_i$ should vanish at illegal configurations. In the realization of NAQS, any spin-1/2 configuration $(s_{\mu_{l}},s_{\mu_{i-1}},\cdots,s_{\mu_1})$ can be encoded by an $l$-digit binary number where $0$ denotes $\downarrow$ and $1$ denotes $\uparrow$. Then $\phi_i$ can be parameterized by neural networks with the input $(s'_{\mu_{i}},s_{\mu_{i-1}},\cdots,s_{\mu_1})$ and the output $\phi_i(s'_{\mu_i}\vert s_{\mu_{i-1}},\cdots,s_{\mu_1})$. In practice, we find having $l$ different conditional wave function for a long chain ($l>20$) is quite clumsy and hard to optimize. So it is helpful to group consecutive spins together. Supposing $l$ can be divided by 4, a convenient strategy is to convert the $l$-digit binary number associated to a spin configuration to its hexadecimal equivalent. For example, a spin configuration $(0,1,1,0,1,1,0,0)$ is converted to ((0110),(1100)). This way the number of conditional wave function is reduced to one fourth of its original number. Let $(h_{\nu_{l/4}},h_{\nu_{l/4-1}},\cdots,h_{\nu_1})$ be the hexadecimal equivalent of $(s_{\mu_{l}},s_{\mu_{i-1}},\cdots,s_{\mu_1})$. The total wave function is given as 
\begin{equation}
        \Upsilon(s_1,\cdots,s_l) = \prod_{i=1}^{l/4} \tilde{\phi_i}(h_{\nu_i}\vert h_{\nu_{i-1}},\cdots,h_{\nu_1})
\end{equation}
satisfying 
\begin{equation}
    \sum_{h'_{\nu_i}\in\{0,1\}^4} \vert\tilde{\phi_i}(h'_{\nu_i}\vert h_{\nu_{i-1}},\cdots,h_{\nu_1})\vert^2
    =1.
\end{equation}

\begin{figure}[tb]
    \centering
    \includegraphics[width=0.9\linewidth]{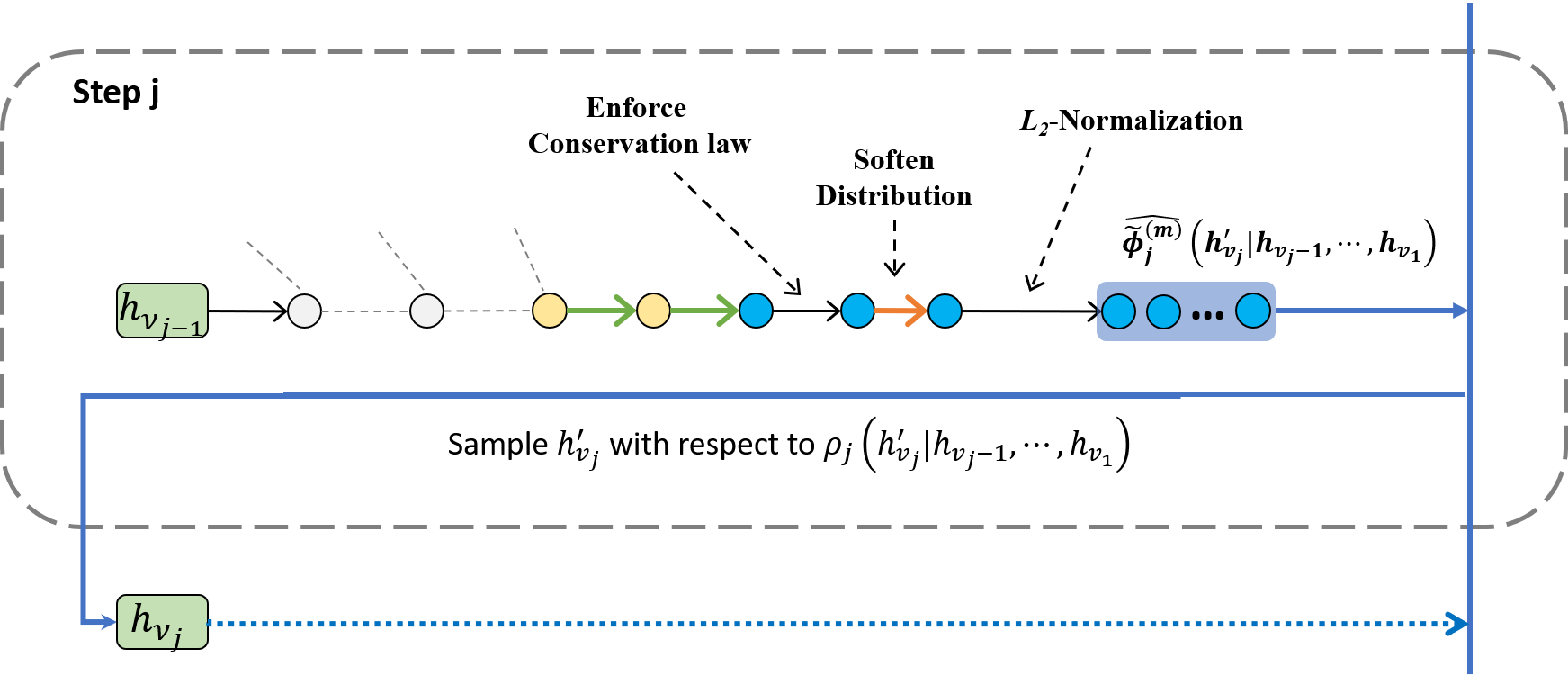}
    \caption{Schematic representation of the direct sampling process (sampling mode) of CGSP-adapted NAQS. The forward propagation is different in the sampling mode in two aspects. First, between enforcing conservation law and enforcing $L_2$-normalization, there is an extra operation devoted to softening the distribution to be sampled. The softening only acts on the (real) amplitude part of the unnormalized wave function through $x\rightarrow x|x|^{\gamma-1}$. The purpose is to prevent mode collapse, i.e. the neural network keeps generating a small set of samples. The resultant normalized conditional wave function is denoted by $\widehat{{\tilde{\phi}_j^m}}$,  to be distinguished from the original ${\tilde{\phi}_j^m}$. The second difference in the sampling mode is that the post-processing procedure is terminated right away after  obtaining $\widehat{{\tilde{\phi}_j^m}}$. }
    \label{fig-sample}
\end{figure}
In CGSP, we need multiple linearly independent wave functions $\{\Upsilon_m(s_1,\cdots,s_l)\}_{m\in[1,M]}$. It will be unnecessarily expensive if each of them is represented with totally independent NAQS. It is wiser allowing them to share part of the parameters. Because non-linearity is applied in each hidden layer of a deep neural network, the sharing of some parameters will not violate the linear independence of the obtained $M$ wave functions. In practice, we let the sharing of parameters happen at the first several hidden layers, which can be understood as a global embedding process. 

Fig.~\ref{fig-naqs} is a schematic representation of the NAQS satisfying these requirements. We name it CGSP-adapted NAQS for ease of reference. A detailed explanation of the forward propagation of CGSP-adapted NAQS  is below the figure.
It is worth mentioning that our design of CGSP-adapted NAQS is inspired by WaveNet~\cite{Oord2016}, where dilated convolution with exponentially increasing dilation size is used to limit the depth of the neural network. We use the same technique in CGSP-adapted NAQS. So the number of convolution layers required by the conditional dependence of NAQS is only $O(\log l)$. 

The carefully designed structure of CGSP-adapted NAQS enables the direct sampling of spin configurations in an efficient parallel manner as illustrated by
Fig.~\ref{fig-sample}. In the sampling mode of CGSP-adapted NAQS,  we will use an auxiliary NAQS $\Upsilon_0(s_1,\cdots,s_l) = \prod_{i=1}^{l/4} \tilde{\phi}_i^{(0)}(h_{\nu_i}\vert h_{\nu_{i-1}},\cdots,h_{\nu_1})$ also satisfying local normalization condition for approximating the initial state $\Psi_0(0)$. When $\Psi_0(0)$ is a simple product state, $\Upsilon_0(s_1,\cdots,s_l)$ can be easily constructed as an exact representation of  $\Psi_0(0)$ and used in both evaluation and sampling mode of CGSP-adapted NAQS. Otherwise, $\Upsilon_0(s_1,\cdots,s_l)$ will only be utilized in the sampling mode.

The whole sampling process consists of $l/4$ steps. For the initial step, $N_B$ empty (all-zero) spin configurations are generated to be placeholders and fed into the neural network. The softened conditional wave function (softening operation explained in the caption of Fig.~\ref{fig-sample})  $\{\widehat{{\tilde{\phi}_1^m}}(h'_{\nu_1})\}_{m\in[1,M]}$ is obtained to sample $h'_{\nu_1}$ with respect to the distribution
\begin{equation}
    \rho_1(h'_{\nu_1}) = \frac{\sum_{m=0}^M w_m\|\widehat{{\tilde{\phi}_1^m}}(h'_{\nu_1})\|^2 }{\sum_{m=0}^Mw_m}
\end{equation}
where the importance weight $w_m$ is suggested by the matrix $A$ in the Eq.~(\ref{theta_cons}) of the main text.
Then the first position of the $N_B$ placeholders are updated accordingly. 

The $j$-th ($1<j\leq l/4$) step of the sampling process is feeding the $N_B$ placeholders back into the neural network and obtaining $\{\widehat{{\tilde{\phi}_j^m}}(h'_{\nu_j}\vert h_{\nu_{j-1}},\cdots,h_{\nu_{1}})\}_{m\in[1,M]}$. Then $h'_{\nu_j}$ is sampled with respect to the distribution 
\begin{widetext}
\begin{equation}
    \rho_j(h'_{\nu_j}\vert h_{\nu_{j-1}},\cdots,h_{\nu_{1}}) = \frac{\sum_{m=0}^M w_m
    \|\widehat{{\tilde{\phi}_j^m}}(h'_{\nu_j}\vert h_{\nu_{j-1}},\cdots,h_{\nu_{1}})\|^2  \prod_{p=1}^{j-1}\|\widehat{{\tilde{\phi}_p^m}}(h_{\nu_p}\vert h_{\nu_{p-1}},\cdots,h_{\nu_{1}})\|^2 
    }{\sum_{m=0}^Mw_m
    \prod_{p=1}^{j-1}
    \|\widehat{{\tilde{\phi}_p^m}}(h_{\nu_p}\vert h_{\nu_{p-1}},\cdots,h_{\nu_{1}})\|^2 }.
\end{equation}
\end{widetext}
It is straightforward to verify that $\rho_j(h'_{\nu_j}\vert h_{\nu_{j-1}},\cdots,h_{\nu_{1}})$ also satisfies the local normalization condition
\begin{equation}
  \sum_{h'_{\nu_j}\in[0,1]^4}  \rho_j(h'_{\nu_j}\vert h_{\nu_{j-1}},\cdots,h_{\nu_{1}})=1.
\end{equation}
Therefore the target probability distribution of the whole sampling process can be expressed  as
\begin{equation}
    P(h'_{\nu_{1}},\cdots, h'_{\nu_{l/4}}) = \prod_{j=1}^{l/4}\rho_j(h'_{\nu_j}\vert h'_{\nu_{j-1}},\cdots,h'_{\nu_{1}}).
\end{equation}
It is easy to see that the normalization condition is automatically satisfied.

There are several {\it ad hoc} parameters to be determined in the sampling mode of CGSP-adapted NAQS. The first one is the real number $0 <\gamma\leq1$  in the softening operation. We find its empirical optimum to be near $0.5$. If $\gamma=1$, this operation is an identity and we find the training of neural network  inefficient and suffering from large local optimum. The second one is the importance weight $w_m$ ($m\in[0,M]$). In our experiments, we let
\begin{equation}
    w_m = \sum_i \left|\frac{\delta_{m0}}{N}+A_{im}-\frac{\sum_{k=0}^{N-1}A_{km}}{N}\right|.
\end{equation}
Besides, there are the permutation $(\mu_1,\cdots,\mu_l)$ of the natural spin order $(1,\cdots, l)$ to be determined. An adequate permutation $\mu$ should minimize ``long range correlation'' in the CGSP-adapted NAQS to control the model complexity. In straightforward terms, $\braket{s_{\mu_i}s_{\mu_j}}_t-\braket{s_{\mu_i}}_t\braket{s_{\mu_j}}_t$ should be small for large $|i-j|$ and the time scale we concern. The design of $\mu$ should also take the symmetry of the initial condition, Hamiltonian and the topology of the lattice into consideration. Empirically, we find the natural spin order is already satisfactory for a 1D chain with open boundary condition. For  periodic boundary condition, the design of $\mu$ relevant to the initial state will require more strategies. 

In summary, the direct sampling algorithm of CGSP-adapted NAQS allows the generation of $N_B$ samples simultaneously through $l$ sequential tailored forward propagation. This is extremely GPU-friendly compared to MCMC algorithms that usually require $O(10^3\sim10^6)$ many sequential forward propagation.  

\section{Technical details of numerical experiments}
The CGSP-adapted NAQS used for our numerical experiments were implemented in PyTorch~\cite{Paszke2019} as introduced in the previous section. For the 32-spin 1D XXZ model and the initial state introduced in the main text, the projected states could all be real functions. So we restricted our CGSP-adapted NAQS for representing real wave functions only.

Our numerical experiments covered two cases: $(M,N)=(32,32)$ and $(M,N)=(32,64)$. Because $M$ was identical in these two simulations, the neural network structure was also identical except the matrix $A=(A_{ij})_{i\in[0,N-1],j\in[0,M]}$ associated to $(M,N)=(32,64)$ has more parameters than the one associated to $(M,N)=(32,32)$. For the forward propagation process in these experiments, the neural network contained $4$ dilated convolution layers ( dilation = $(1,1,2,4)$, kernel size = $(2,2,2,2)$, out channel = $(128,128,128,128)$ ). The outputs of the last three convolution layers were concatenated through the channel dimension and rescaled by a $1\times1$ convolution layer. This completed the ``embedding'' stage in Fig.~\ref{fig-naqs} and yielded a 3D tensor of size $(l/4 ,N_B, 384)$ where $l/4$ corresponded to the number of vertical nodes in Fig.~\ref{fig-naqs}. The next stage ``projection'' had two sparsely connected linear layers. The first linear layer consisted of $l/4$ small fully connected layers operating on each nodes independently, yielding a 3D tensor of size $(l/4 ,N_B, 64M)$. The result was reshaped into a 4D tensor of size $(M, l/4 ,N_B, 64)$ and fed into the second linear layer consisting of $Ml/4$ small fully connected layers assigned to the first two dimensions, yielding a 4D tensor of size $(M, l/4 ,N_B, 16)$. The output was reshaped into size $(M, N_B, l/4, 16)$ which completed  the ``projection'' stage. The last stage ``post-processing''  had been described in the previous section as well as in the main text.
\begin{figure}[t]
    \centering
    \includegraphics[width=\linewidth]{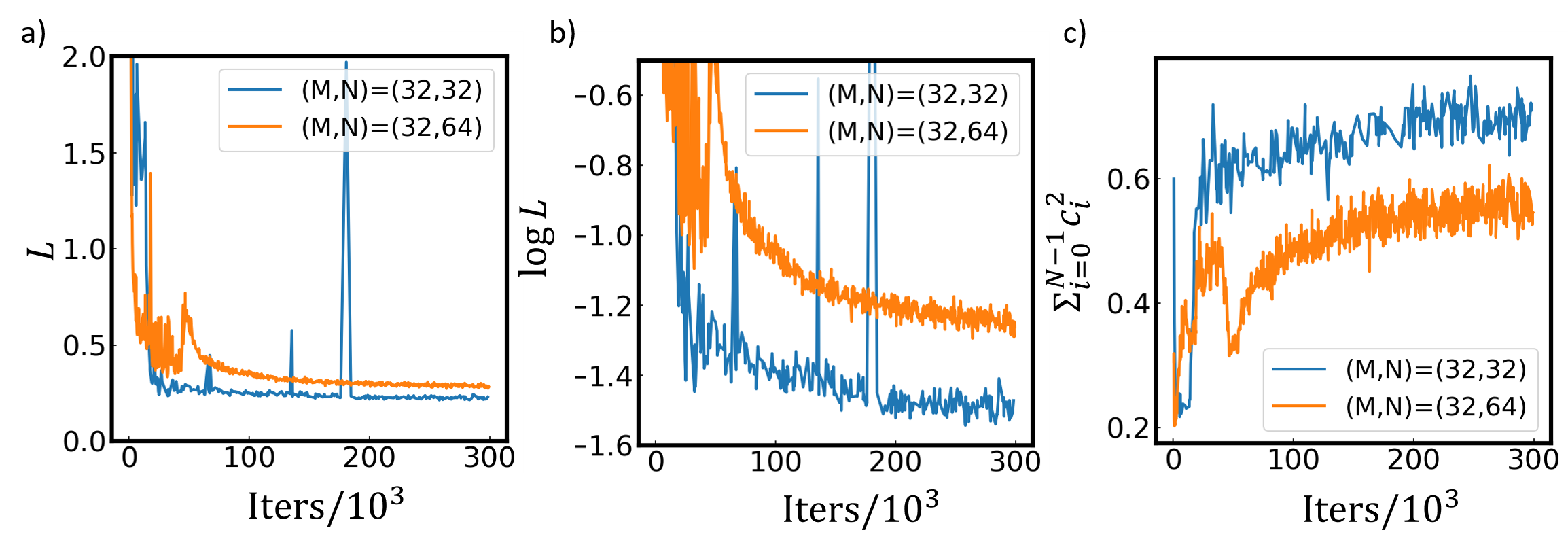}
    \caption{(a) Estimated loss $L$ versus iterations. (b)  Estimated $\log L$ versus iterations. (c) Estimated $\sum_{i=0}^{N-1}c_i^2$ versus iterations. }
    \label{fig-train}
\end{figure}
For 32-spin 1D XXZ model, the dimension of the underlying Hilbert space is about $6\times10^8$ within the zero total $S^z$ sector. The number of trainable parameters in both experiments are about $7\times 10^6$. For about $1\%$ of the Hilbert space complexity, our CGSP-adapted NAQS demonstrated its parameter sharing strategy very efficient. 
The training part of the two experiments was accomplished by ADAM~\cite{kingma2014adam}, a first order gradient descent optimizer with adaptive learning rate for each parameters. We didn't rule out the possibility that second order optimization methods could be more efficient for CGSP tasks. For both experiments, the total number of stochastic samples for each update (iteration) was $4000$ and the learning rate was fixed to $1\times10^{-3}$. We didn't find a learning rate decay improving the convergence. We plotted the training curve in Fig.~\ref{fig-train}. For $(M,N)=(32,32)$, the wall clock time 
for $10^5$ iterations trained with 2 NVIDIA
Tesla V100 GPUs was about 6 hours. For $(M,N)=(32,64)$, the wall clock time 
for $10^5$ iterations trained with 4 NVIDIA
Tesla V100 GPUs was about 5 hours.

\section{A parallel framework for CGSP-breakdown}

\begin{figure}[t]
    \centering
    \includegraphics[width=0.8\linewidth]{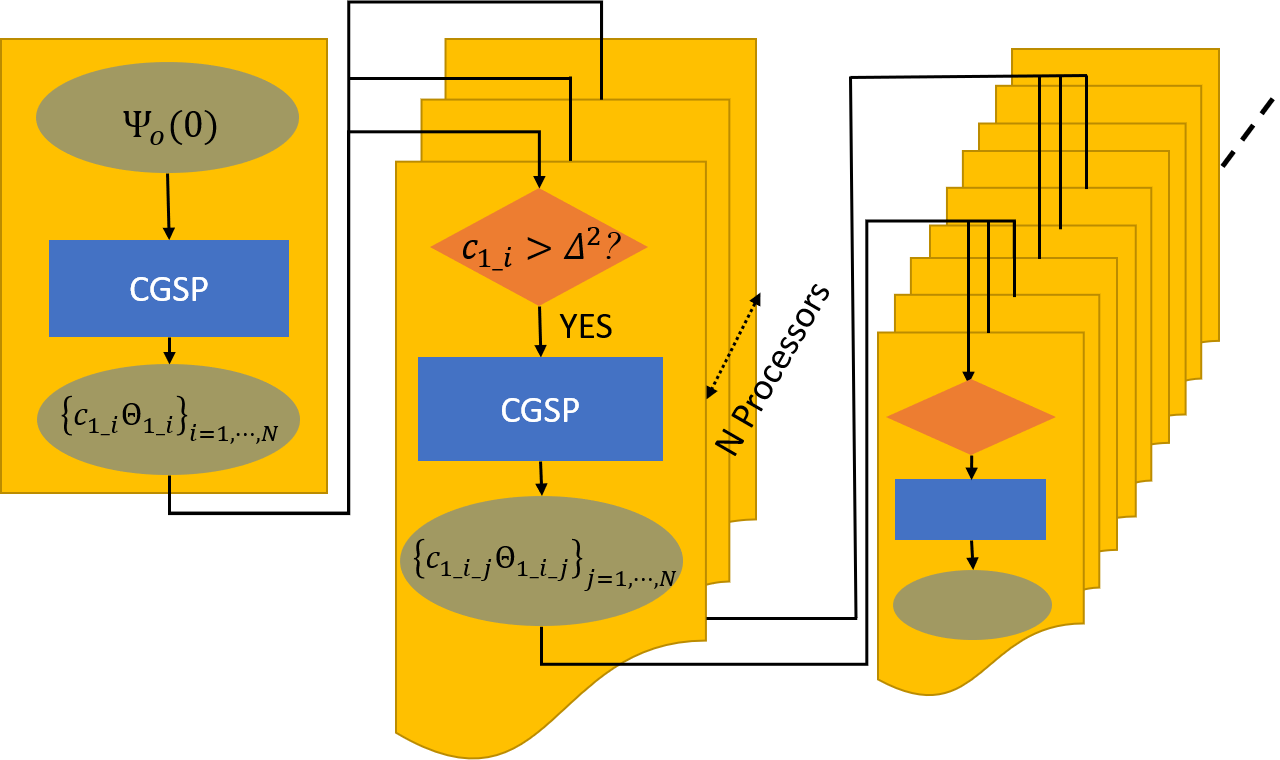}
    \caption{A parallel scheme for CGSP-breakdown. The label of each states implies a tree structure. The label $\iota=o$ denotes the root node, $\iota=\text{1\_i}$ denotes the $i$-th child of the root node and $\iota=\text{1\_i\_j}$ denotes the $j$-th child of $\iota=\text{1\_i}$.    }
    \label{pll}
\end{figure}

Training a large neural network with a complicated loss function can be numerically unstable and troubled by local optimum.
So we propose a simple parallel framework for breaking down CGSP into hierarchically-organized sub-tasks. A flowchart of its realization is shown in Fig.~\ref{pll}, where the whole CGSP process is divided into several layers. An initial CGSP of the initial state $\Psi_o(0)$ is carried out in one processor with affordable $M$ and $N$. Then the  $N$ projected states 
whose  amplitude is above certain threshold $\Delta$ are sent to different processors for the next-layer CGSP. 
Notice the second-step CGSPs are independent and naturally parallel. This procedure can be repeated for higher resolution of the spectrum if satisfactory convergence is achieved in each step. At the end, one obtains a family of neural network quantum states organized in a tree structure.   

To recover the unitary quantum dynamics, the energy expectation $\lambda_\iota $ of leaf state labeled by $\iota$ should be computed for all leaf nodes. The approximation to $\Psi_o(t)$ thus becomes
\begin{equation}
 \varphi_o(t) = \sum_{\iota\in \text{leaf nodes}} c_\iota e^{-i\lambda_\iota t}\Theta_\iota.  
\end{equation}

Note that the numerical experiments presented in the main text does not utilize this framework.

\section{CGSP-initialized TDVP simulation}
Let $H(t)$ be a slowly-varying Hamiltonian that the spectral norm $\|dH(t)/dt\|_{s}$ is bounded by $B>0$. Let $\Psi_o(t)$ be the pure state evolving with $H(t)$.

Suppose the initial state $\Psi_o(0)$ has already found its CGSP representation $\Psi_o(0)=\sum_{i=0}^{N-1}c_i \Theta_i$ with $\lambda_i = \frac{\braket{\Theta_i\vert H(0)\vert \Theta_i}}{\braket{\Theta_i\vert \Theta_i}}$. By allowing the parameters of the neural networks to be time-dependent and identifying $\Theta_i$ as $\Theta_i(t=0)$,  a new ansatz can be defined with $\eta_i(t) = \Theta_i(t)e^{-i\lambda_i t}$. If $\eta_i(t)$ evolves under the Schr\"odinger equation exactly, then $\sum_{i=0}^{N-1} c_i\eta_i(t)$ is identical to $\Psi_o(t)$. The time-dependent variational principle for $\eta_i(t)$ writes:   
\begin{equation}
    \delta \int_0^t \| i\frac{d\eta_i(s)}{d s} - H(s)\eta_i(s) \|^2 ds = 0,
\end{equation}
which can be translated into
\begin{equation}
    \mathop{\text{minimize}}  \|i\frac{d\Theta_i(t)}{dt} - (H(t) - \lambda_i) \Theta_i(t)\|.
\end{equation}
Considering the Hamiltonian is slowly varying, there is 
\begin{equation}\label{tdvpbd}
    \|\frac{d\Theta_i(t)}{dt}\| < Bt + \|(H(0) - \lambda_i)\Theta_i(t)\|
\end{equation}

If  CGSP is successful, one expects $\|(H(0) - \lambda_i)\Theta_i(t)\| \ll 1$ for $t$ small enough. 
Eq.~(\ref{tdvpbd}) suggests how CGSP may help with TDVP-based simulation. For the plain TDVP approach, the variation of the ansatz $||\frac{d\Psi(t)}{d t}||$ is bounded by $||H(t)||_s$, which usually grows linearly with system size for lattice models. This means the differentiable manifold that the neural network should parameterize grows rapidly for ergodic dynamics. However, with CGSP-initialized TDVP simulation, the desired expressive power of the neural network grows much slower due to the constraint Eq.~(\ref{tdvpbd}).

\bibliographystyle{apsrev4-2}
\bibliography{cgsp}

\begin{thebibliography}{52}%
\makeatletter
\providecommand \@ifxundefined [1]{%
 \@ifx{#1\undefined}
}%
\providecommand \@ifnum [1]{%
 \ifnum #1\expandafter \@firstoftwo
 \else \expandafter \@secondoftwo
 \fi
}%
\providecommand \@ifx [1]{%
 \ifx #1\expandafter \@firstoftwo
 \else \expandafter \@secondoftwo
 \fi
}%
\providecommand \natexlab [1]{#1}%
\providecommand \enquote  [1]{``#1''}%
\providecommand \bibnamefont  [1]{#1}%
\providecommand \bibfnamefont [1]{#1}%
\providecommand \citenamefont [1]{#1}%
\providecommand \href@noop [0]{\@secondoftwo}%
\providecommand \href [0]{\begingroup \@sanitize@url \@href}%
\providecommand \@href[1]{\@@startlink{#1}\@@href}%
\providecommand \@@href[1]{\endgroup#1\@@endlink}%
\providecommand \@sanitize@url [0]{\catcode `\\12\catcode `\$12\catcode
  `\&12\catcode `\#12\catcode `\^12\catcode `\_12\catcode `\%12\relax}%
\providecommand \@@startlink[1]{}%
\providecommand \@@endlink[0]{}%
\providecommand \url  [0]{\begingroup\@sanitize@url \@url }%
\providecommand \@url [1]{\endgroup\@href {#1}{\urlprefix }}%
\providecommand \urlprefix  [0]{URL }%
\providecommand \Eprint [0]{\href }%
\providecommand \doibase [0]{https://doi.org/}%
\providecommand \selectlanguage [0]{\@gobble}%
\providecommand \bibinfo  [0]{\@secondoftwo}%
\providecommand \bibfield  [0]{\@secondoftwo}%
\providecommand \translation [1]{[#1]}%
\providecommand \BibitemOpen [0]{}%
\providecommand \bibitemStop [0]{}%
\providecommand \bibitemNoStop [0]{.\EOS\space}%
\providecommand \EOS [0]{\spacefactor3000\relax}%
\providecommand \BibitemShut  [1]{\csname bibitem#1\endcsname}%
\let\auto@bib@innerbib\@empty
\bibitem [{\citenamefont {Kaufman}\ \emph {et~al.}(2012)\citenamefont
  {Kaufman}, \citenamefont {Lester},\ and\ \citenamefont
  {Regal}}]{Kaufman2012}%
  \BibitemOpen
  \bibfield  {author} {\bibinfo {author} {\bibfnamefont {A.~M.}\ \bibnamefont
  {Kaufman}}, \bibinfo {author} {\bibfnamefont {B.~J.}\ \bibnamefont
  {Lester}},\ and\ \bibinfo {author} {\bibfnamefont {C.~A.}\ \bibnamefont
  {Regal}},\ }\href {https://doi.org/10.1103/PhysRevX.2.041014} {\bibfield
  {journal} {\bibinfo  {journal} {Physical Review X}\ }\textbf {\bibinfo
  {volume} {2}},\ \bibinfo {pages} {041014} (\bibinfo {year} {2012})},\ \Eprint
  {https://arxiv.org/abs/1209.2087} {arXiv:1209.2087} \BibitemShut {NoStop}%
\bibitem [{\citenamefont {Phillips}(1998)}]{Phillips1998}%
  \BibitemOpen
  \bibfield  {author} {\bibinfo {author} {\bibfnamefont {W.~D.}\ \bibnamefont
  {Phillips}},\ }\href {https://doi.org/10.1103/revmodphys.70.721} {\bibfield
  {journal} {\bibinfo  {journal} {Reviews of Modern Physics}\ }\textbf
  {\bibinfo {volume} {70}},\ \bibinfo {pages} {721} (\bibinfo {year}
  {1998})}\BibitemShut {NoStop}%
\bibitem [{\citenamefont {Weiss}\ and\ \citenamefont
  {Saffman}(2017)}]{Weiss2017}%
  \BibitemOpen
  \bibfield  {author} {\bibinfo {author} {\bibfnamefont {D.~S.}\ \bibnamefont
  {Weiss}}\ and\ \bibinfo {author} {\bibfnamefont {M.}~\bibnamefont
  {Saffman}},\ }\href {https://doi.org/10.1063/PT.3.3626} {\bibfield  {journal}
  {\bibinfo  {journal} {Physics Today}\ }\textbf {\bibinfo {volume} {70}},\
  \bibinfo {pages} {44} (\bibinfo {year} {2017})}\BibitemShut {NoStop}%
\bibitem [{\citenamefont {Spring}\ \emph {et~al.}(2013)\citenamefont {Spring},
  \citenamefont {Metcalf}, \citenamefont {Humphreys}, \citenamefont
  {Kolthammer}, \citenamefont {Jin}, \citenamefont {Barbieri}, \citenamefont
  {Datta}, \citenamefont {Thomas-Peter}, \citenamefont {Langford},
  \citenamefont {Kundys} \emph {et~al.}}]{Spring2013}%
  \BibitemOpen
  \bibfield  {author} {\bibinfo {author} {\bibfnamefont {J.~B.}\ \bibnamefont
  {Spring}}, \bibinfo {author} {\bibfnamefont {B.~J.}\ \bibnamefont {Metcalf}},
  \bibinfo {author} {\bibfnamefont {P.~C.}\ \bibnamefont {Humphreys}}, \bibinfo
  {author} {\bibfnamefont {W.~S.}\ \bibnamefont {Kolthammer}}, \bibinfo
  {author} {\bibfnamefont {X.-M.}\ \bibnamefont {Jin}}, \bibinfo {author}
  {\bibfnamefont {M.}~\bibnamefont {Barbieri}}, \bibinfo {author}
  {\bibfnamefont {A.}~\bibnamefont {Datta}}, \bibinfo {author} {\bibfnamefont
  {N.}~\bibnamefont {Thomas-Peter}}, \bibinfo {author} {\bibfnamefont {N.~K.}\
  \bibnamefont {Langford}}, \bibinfo {author} {\bibfnamefont {D.}~\bibnamefont
  {Kundys}}, \emph {et~al.},\ }\href@noop {} {\bibfield  {journal} {\bibinfo
  {journal} {Science}\ }\textbf {\bibinfo {volume} {339}},\ \bibinfo {pages}
  {798} (\bibinfo {year} {2013})}\BibitemShut {NoStop}%
\bibitem [{\citenamefont {Vandersypen}\ and\ \citenamefont
  {Chuang}(2004)}]{Vandersypen2004}%
  \BibitemOpen
  \bibfield  {author} {\bibinfo {author} {\bibfnamefont {L.~M.}\ \bibnamefont
  {Vandersypen}}\ and\ \bibinfo {author} {\bibfnamefont {I.~L.}\ \bibnamefont
  {Chuang}},\ }\href {https://doi.org/10.1103/RevModPhys.76.1037} {\bibinfo
  {title} {{NMR techniques for quantum control and computation}}} (\bibinfo
  {year} {2004}),\ \Eprint {https://arxiv.org/abs/0404064} {arXiv:0404064
  [quant-ph]} \BibitemShut {NoStop}%
\bibitem [{\citenamefont {Schmitz}\ \emph {et~al.}(2009)\citenamefont
  {Schmitz}, \citenamefont {Matjeschk}, \citenamefont {Schneider},
  \citenamefont {Glueckert}, \citenamefont {Enderlein}, \citenamefont {Huber},\
  and\ \citenamefont {Schaetz}}]{Schmitz2009}%
  \BibitemOpen
  \bibfield  {author} {\bibinfo {author} {\bibfnamefont {H.}~\bibnamefont
  {Schmitz}}, \bibinfo {author} {\bibfnamefont {R.}~\bibnamefont {Matjeschk}},
  \bibinfo {author} {\bibfnamefont {C.}~\bibnamefont {Schneider}}, \bibinfo
  {author} {\bibfnamefont {J.}~\bibnamefont {Glueckert}}, \bibinfo {author}
  {\bibfnamefont {M.}~\bibnamefont {Enderlein}}, \bibinfo {author}
  {\bibfnamefont {T.}~\bibnamefont {Huber}},\ and\ \bibinfo {author}
  {\bibfnamefont {T.}~\bibnamefont {Schaetz}},\ }\href
  {https://doi.org/10.1103/PhysRevLett.103.090504} {\bibfield  {journal}
  {\bibinfo  {journal} {Physical Review Letters}\ }\textbf {\bibinfo {volume}
  {103}},\ \bibinfo {pages} {090504} (\bibinfo {year} {2009})},\ \Eprint
  {https://arxiv.org/abs/0904.4214} {arXiv:0904.4214} \BibitemShut {NoStop}%
\bibitem [{\citenamefont {Zhang}\ \emph
  {et~al.}(2017{\natexlab{a}})\citenamefont {Zhang}, \citenamefont {Pagano},
  \citenamefont {Hess}, \citenamefont {Kyprianidis}, \citenamefont {Becker},
  \citenamefont {Kaplan}, \citenamefont {Gorshkov}, \citenamefont {Gong},\ and\
  \citenamefont {Monroe}}]{Zhang2017a}%
  \BibitemOpen
  \bibfield  {author} {\bibinfo {author} {\bibfnamefont {J.}~\bibnamefont
  {Zhang}}, \bibinfo {author} {\bibfnamefont {G.}~\bibnamefont {Pagano}},
  \bibinfo {author} {\bibfnamefont {P.~W.}\ \bibnamefont {Hess}}, \bibinfo
  {author} {\bibfnamefont {A.}~\bibnamefont {Kyprianidis}}, \bibinfo {author}
  {\bibfnamefont {P.}~\bibnamefont {Becker}}, \bibinfo {author} {\bibfnamefont
  {H.}~\bibnamefont {Kaplan}}, \bibinfo {author} {\bibfnamefont {A.~V.}\
  \bibnamefont {Gorshkov}}, \bibinfo {author} {\bibfnamefont {Z.~X.}\
  \bibnamefont {Gong}},\ and\ \bibinfo {author} {\bibfnamefont
  {C.}~\bibnamefont {Monroe}},\ }\href {https://doi.org/10.1038/nature24654}
  {\bibfield  {journal} {\bibinfo  {journal} {Nature}\ }\textbf {\bibinfo
  {volume} {551}},\ \bibinfo {pages} {601} (\bibinfo {year}
  {2017}{\natexlab{a}})},\ \Eprint {https://arxiv.org/abs/1708.01044}
  {arXiv:1708.01044} \BibitemShut {NoStop}%
\bibitem [{\citenamefont {Zhang}\ \emph
  {et~al.}(2017{\natexlab{b}})\citenamefont {Zhang}, \citenamefont {Hess},
  \citenamefont {Kyprianidis}, \citenamefont {Becker}, \citenamefont {Lee},
  \citenamefont {Smith}, \citenamefont {Pagano}, \citenamefont {Potirniche},
  \citenamefont {Potter}, \citenamefont {Vishwanath}, \citenamefont {Yao},\
  and\ \citenamefont {Monroe}}]{Zhang2017}%
  \BibitemOpen
  \bibfield  {author} {\bibinfo {author} {\bibfnamefont {J.}~\bibnamefont
  {Zhang}}, \bibinfo {author} {\bibfnamefont {P.~W.}\ \bibnamefont {Hess}},
  \bibinfo {author} {\bibfnamefont {A.}~\bibnamefont {Kyprianidis}}, \bibinfo
  {author} {\bibfnamefont {P.}~\bibnamefont {Becker}}, \bibinfo {author}
  {\bibfnamefont {A.}~\bibnamefont {Lee}}, \bibinfo {author} {\bibfnamefont
  {J.}~\bibnamefont {Smith}}, \bibinfo {author} {\bibfnamefont
  {G.}~\bibnamefont {Pagano}}, \bibinfo {author} {\bibfnamefont {I.~D.}\
  \bibnamefont {Potirniche}}, \bibinfo {author} {\bibfnamefont {A.~C.}\
  \bibnamefont {Potter}}, \bibinfo {author} {\bibfnamefont {A.}~\bibnamefont
  {Vishwanath}}, \bibinfo {author} {\bibfnamefont {N.~Y.}\ \bibnamefont
  {Yao}},\ and\ \bibinfo {author} {\bibfnamefont {C.}~\bibnamefont {Monroe}},\
  }\href {https://doi.org/10.1038/nature21413} {\bibfield  {journal} {\bibinfo
  {journal} {Nature}\ }\textbf {\bibinfo {volume} {543}},\ \bibinfo {pages}
  {217} (\bibinfo {year} {2017}{\natexlab{b}})},\ \Eprint
  {https://arxiv.org/abs/1609.08684} {arXiv:1609.08684} \BibitemShut {NoStop}%
\bibitem [{\citenamefont {Smith}\ \emph {et~al.}(2016)\citenamefont {Smith},
  \citenamefont {Lee}, \citenamefont {Richerme}, \citenamefont {Neyenhuis},
  \citenamefont {Hess}, \citenamefont {Hauke}, \citenamefont {Heyl},
  \citenamefont {Huse},\ and\ \citenamefont {Monroe}}]{Smith2016}%
  \BibitemOpen
  \bibfield  {author} {\bibinfo {author} {\bibfnamefont {J.}~\bibnamefont
  {Smith}}, \bibinfo {author} {\bibfnamefont {A.}~\bibnamefont {Lee}}, \bibinfo
  {author} {\bibfnamefont {P.}~\bibnamefont {Richerme}}, \bibinfo {author}
  {\bibfnamefont {B.}~\bibnamefont {Neyenhuis}}, \bibinfo {author}
  {\bibfnamefont {P.~W.}\ \bibnamefont {Hess}}, \bibinfo {author}
  {\bibfnamefont {P.}~\bibnamefont {Hauke}}, \bibinfo {author} {\bibfnamefont
  {M.}~\bibnamefont {Heyl}}, \bibinfo {author} {\bibfnamefont {D.~A.}\
  \bibnamefont {Huse}},\ and\ \bibinfo {author} {\bibfnamefont
  {C.}~\bibnamefont {Monroe}},\ }\href {https://doi.org/10.1038/nphys3783}
  {\bibfield  {journal} {\bibinfo  {journal} {Nature Physics}\ }\textbf
  {\bibinfo {volume} {12}},\ \bibinfo {pages} {907} (\bibinfo {year} {2016})},\
  \Eprint {https://arxiv.org/abs/1508.07026} {arXiv:1508.07026} \BibitemShut
  {NoStop}%
\bibitem [{\citenamefont {Gring}\ \emph {et~al.}(2012)\citenamefont {Gring},
  \citenamefont {Kuhnert}, \citenamefont {Langen}, \citenamefont {Kitagawa},
  \citenamefont {Rauer}, \citenamefont {Schreitl}, \citenamefont {Mazets},
  \citenamefont {Smith}, \citenamefont {Demler},\ and\ \citenamefont
  {Schmiedmayer}}]{Gring2012}%
  \BibitemOpen
  \bibfield  {author} {\bibinfo {author} {\bibfnamefont {M.}~\bibnamefont
  {Gring}}, \bibinfo {author} {\bibfnamefont {M.}~\bibnamefont {Kuhnert}},
  \bibinfo {author} {\bibfnamefont {T.}~\bibnamefont {Langen}}, \bibinfo
  {author} {\bibfnamefont {T.}~\bibnamefont {Kitagawa}}, \bibinfo {author}
  {\bibfnamefont {B.}~\bibnamefont {Rauer}}, \bibinfo {author} {\bibfnamefont
  {M.}~\bibnamefont {Schreitl}}, \bibinfo {author} {\bibfnamefont
  {I.}~\bibnamefont {Mazets}}, \bibinfo {author} {\bibfnamefont {D.~A.}\
  \bibnamefont {Smith}}, \bibinfo {author} {\bibfnamefont {E.}~\bibnamefont
  {Demler}},\ and\ \bibinfo {author} {\bibfnamefont {J.}~\bibnamefont
  {Schmiedmayer}},\ }\href@noop {} {\bibfield  {journal} {\bibinfo  {journal}
  {Science}\ }\textbf {\bibinfo {volume} {337}},\ \bibinfo {pages} {1318}
  (\bibinfo {year} {2012})}\BibitemShut {NoStop}%
\bibitem [{\citenamefont {Cai}\ \emph {et~al.}(2013)\citenamefont {Cai},
  \citenamefont {Weedbrook}, \citenamefont {Su}, \citenamefont {Chen},
  \citenamefont {Gu}, \citenamefont {Zhu}, \citenamefont {Li}, \citenamefont
  {Liu}, \citenamefont {Lu},\ and\ \citenamefont {Pan}}]{Cai2013}%
  \BibitemOpen
  \bibfield  {author} {\bibinfo {author} {\bibfnamefont {X.-D.}\ \bibnamefont
  {Cai}}, \bibinfo {author} {\bibfnamefont {C.}~\bibnamefont {Weedbrook}},
  \bibinfo {author} {\bibfnamefont {Z.-E.}\ \bibnamefont {Su}}, \bibinfo
  {author} {\bibfnamefont {M.-C.}\ \bibnamefont {Chen}}, \bibinfo {author}
  {\bibfnamefont {M.}~\bibnamefont {Gu}}, \bibinfo {author} {\bibfnamefont
  {M.-J.}\ \bibnamefont {Zhu}}, \bibinfo {author} {\bibfnamefont
  {L.}~\bibnamefont {Li}}, \bibinfo {author} {\bibfnamefont {N.-L.}\
  \bibnamefont {Liu}}, \bibinfo {author} {\bibfnamefont {C.-Y.}\ \bibnamefont
  {Lu}},\ and\ \bibinfo {author} {\bibfnamefont {J.-W.}\ \bibnamefont {Pan}},\
  }\href@noop {} {\bibfield  {journal} {\bibinfo  {journal} {Physical review
  letters}\ }\textbf {\bibinfo {volume} {110}},\ \bibinfo {pages} {230501}
  (\bibinfo {year} {2013})}\BibitemShut {NoStop}%
\bibitem [{\citenamefont {Houck}\ \emph {et~al.}(2012)\citenamefont {Houck},
  \citenamefont {T{\"{u}}reci},\ and\ \citenamefont {Koch}}]{Houck2012}%
  \BibitemOpen
  \bibfield  {author} {\bibinfo {author} {\bibfnamefont {A.~A.}\ \bibnamefont
  {Houck}}, \bibinfo {author} {\bibfnamefont {H.~E.}\ \bibnamefont
  {T{\"{u}}reci}},\ and\ \bibinfo {author} {\bibfnamefont {J.}~\bibnamefont
  {Koch}},\ }\href {https://doi.org/10.1038/nphys2251} {\bibfield  {journal}
  {\bibinfo  {journal} {Nature Physics}\ }\textbf {\bibinfo {volume} {8}},\
  \bibinfo {pages} {292} (\bibinfo {year} {2012})}\BibitemShut {NoStop}%
\bibitem [{\citenamefont {Harty}\ \emph {et~al.}(2014)\citenamefont {Harty},
  \citenamefont {Allcock}, \citenamefont {Ballance}, \citenamefont {Guidoni},
  \citenamefont {Janacek}, \citenamefont {Linke}, \citenamefont {Stacey},\ and\
  \citenamefont {Lucas}}]{Harty2014}%
  \BibitemOpen
  \bibfield  {author} {\bibinfo {author} {\bibfnamefont {T.~P.}\ \bibnamefont
  {Harty}}, \bibinfo {author} {\bibfnamefont {D.~T.}\ \bibnamefont {Allcock}},
  \bibinfo {author} {\bibfnamefont {C.~J.}\ \bibnamefont {Ballance}}, \bibinfo
  {author} {\bibfnamefont {L.}~\bibnamefont {Guidoni}}, \bibinfo {author}
  {\bibfnamefont {H.~A.}\ \bibnamefont {Janacek}}, \bibinfo {author}
  {\bibfnamefont {N.~M.}\ \bibnamefont {Linke}}, \bibinfo {author}
  {\bibfnamefont {D.~N.}\ \bibnamefont {Stacey}},\ and\ \bibinfo {author}
  {\bibfnamefont {D.~M.}\ \bibnamefont {Lucas}},\ }\href
  {https://doi.org/10.1103/PhysRevLett.113.220501} {\bibfield  {journal}
  {\bibinfo  {journal} {Physical Review Letters}\ }\textbf {\bibinfo {volume}
  {113}},\ \bibinfo {pages} {220501} (\bibinfo {year} {2014})},\ \Eprint
  {https://arxiv.org/abs/1403.1524} {arXiv:1403.1524} \BibitemShut {NoStop}%
\bibitem [{\citenamefont {Arute}\ \emph {et~al.}(2019)\citenamefont {Arute},
  \citenamefont {Arya}, \citenamefont {Babbush}, \citenamefont {Bacon},
  \citenamefont {Bardin}, \citenamefont {Barends}, \citenamefont {Biswas},
  \citenamefont {Boixo}, \citenamefont {Brandao}, \citenamefont {Buell} \emph
  {et~al.}}]{Arute2019}%
  \BibitemOpen
  \bibfield  {author} {\bibinfo {author} {\bibfnamefont {F.}~\bibnamefont
  {Arute}}, \bibinfo {author} {\bibfnamefont {K.}~\bibnamefont {Arya}},
  \bibinfo {author} {\bibfnamefont {R.}~\bibnamefont {Babbush}}, \bibinfo
  {author} {\bibfnamefont {D.}~\bibnamefont {Bacon}}, \bibinfo {author}
  {\bibfnamefont {J.~C.}\ \bibnamefont {Bardin}}, \bibinfo {author}
  {\bibfnamefont {R.}~\bibnamefont {Barends}}, \bibinfo {author} {\bibfnamefont
  {R.}~\bibnamefont {Biswas}}, \bibinfo {author} {\bibfnamefont
  {S.}~\bibnamefont {Boixo}}, \bibinfo {author} {\bibfnamefont {F.~G.}\
  \bibnamefont {Brandao}}, \bibinfo {author} {\bibfnamefont {D.~A.}\
  \bibnamefont {Buell}}, \emph {et~al.},\ }\href@noop {} {\bibfield  {journal}
  {\bibinfo  {journal} {Nature}\ }\textbf {\bibinfo {volume} {574}},\ \bibinfo
  {pages} {505} (\bibinfo {year} {2019})}\BibitemShut {NoStop}%
\bibitem [{\citenamefont {Pan}\ \emph {et~al.}(2012)\citenamefont {Pan},
  \citenamefont {Chen}, \citenamefont {Lu}, \citenamefont {Weinfurter},
  \citenamefont {Zeilinger},\ and\ \citenamefont {Zukowski}}]{Pan2012}%
  \BibitemOpen
  \bibfield  {author} {\bibinfo {author} {\bibfnamefont {J.~W.}\ \bibnamefont
  {Pan}}, \bibinfo {author} {\bibfnamefont {Z.~B.}\ \bibnamefont {Chen}},
  \bibinfo {author} {\bibfnamefont {C.~Y.}\ \bibnamefont {Lu}}, \bibinfo
  {author} {\bibfnamefont {H.}~\bibnamefont {Weinfurter}}, \bibinfo {author}
  {\bibfnamefont {A.}~\bibnamefont {Zeilinger}},\ and\ \bibinfo {author}
  {\bibfnamefont {M.}~\bibnamefont {Zukowski}},\ }\href
  {https://doi.org/10.1103/RevModPhys.84.777} {\bibfield  {journal} {\bibinfo
  {journal} {Reviews of Modern Physics}\ }\textbf {\bibinfo {volume} {84}},\
  \bibinfo {pages} {777} (\bibinfo {year} {2012})},\ \Eprint
  {https://arxiv.org/abs/0805.2853} {arXiv:0805.2853} \BibitemShut {NoStop}%
\bibitem [{\citenamefont {Wahl}\ \emph
  {et~al.}(2017{\natexlab{a}})\citenamefont {Wahl}, \citenamefont {Pal},\ and\
  \citenamefont {Simon}}]{Wahl}%
  \BibitemOpen
  \bibfield  {author} {\bibinfo {author} {\bibfnamefont {T.~B.}\ \bibnamefont
  {Wahl}}, \bibinfo {author} {\bibfnamefont {A.}~\bibnamefont {Pal}},\ and\
  \bibinfo {author} {\bibfnamefont {S.~H.}\ \bibnamefont {Simon}},\ }\href@noop
  {} {\bibfield  {journal} {\bibinfo  {journal} {Physical Review X}\ }\textbf
  {\bibinfo {volume} {7}},\ \bibinfo {pages} {021018} (\bibinfo {year}
  {2017}{\natexlab{a}})}\BibitemShut {NoStop}%
\bibitem [{\citenamefont {Devakul}\ and\ \citenamefont
  {Singh}(2015)}]{Devakul2015}%
  \BibitemOpen
  \bibfield  {author} {\bibinfo {author} {\bibfnamefont {T.}~\bibnamefont
  {Devakul}}\ and\ \bibinfo {author} {\bibfnamefont {R.~R.}\ \bibnamefont
  {Singh}},\ }\href@noop {} {\bibfield  {journal} {\bibinfo  {journal}
  {Physical review letters}\ }\textbf {\bibinfo {volume} {115}},\ \bibinfo
  {pages} {187201} (\bibinfo {year} {2015})}\BibitemShut {NoStop}%
\bibitem [{\citenamefont {Schr{\"{o}}der}\ \emph {et~al.}(2019)\citenamefont
  {Schr{\"{o}}der}, \citenamefont {Turban}, \citenamefont {Musser},
  \citenamefont {Hine},\ and\ \citenamefont {Chin}}]{Schroder2019}%
  \BibitemOpen
  \bibfield  {author} {\bibinfo {author} {\bibfnamefont {F.~A.}\ \bibnamefont
  {Schr{\"{o}}der}}, \bibinfo {author} {\bibfnamefont {D.~H.}\ \bibnamefont
  {Turban}}, \bibinfo {author} {\bibfnamefont {A.~J.}\ \bibnamefont {Musser}},
  \bibinfo {author} {\bibfnamefont {N.~D.}\ \bibnamefont {Hine}},\ and\
  \bibinfo {author} {\bibfnamefont {A.~W.}\ \bibnamefont {Chin}},\ }\href
  {https://doi.org/10.1038/s41467-019-09039-7} {\bibfield  {journal} {\bibinfo
  {journal} {Nature Communications}\ }\textbf {\bibinfo {volume} {10}},\
  \bibinfo {pages} {1} (\bibinfo {year} {2019})}\BibitemShut {NoStop}%
\bibitem [{\citenamefont {Khasseh}\ \emph {et~al.}(2020)\citenamefont
  {Khasseh}, \citenamefont {Russomanno}, \citenamefont {Schmitt}, \citenamefont
  {Heyl},\ and\ \citenamefont {Fazio}}]{Khasseh2020}%
  \BibitemOpen
  \bibfield  {author} {\bibinfo {author} {\bibfnamefont {R.}~\bibnamefont
  {Khasseh}}, \bibinfo {author} {\bibfnamefont {A.}~\bibnamefont {Russomanno}},
  \bibinfo {author} {\bibfnamefont {M.}~\bibnamefont {Schmitt}}, \bibinfo
  {author} {\bibfnamefont {M.}~\bibnamefont {Heyl}},\ and\ \bibinfo {author}
  {\bibfnamefont {R.}~\bibnamefont {Fazio}},\ }\href
  {https://doi.org/10.1103/PhysRevB.102.014303} {\bibfield  {journal} {\bibinfo
   {journal} {Physical Review B}\ }\textbf {\bibinfo {volume} {102}},\ \bibinfo
  {pages} {014303} (\bibinfo {year} {2020})}\BibitemShut {NoStop}%
\bibitem [{\citenamefont {{Del Pino}}\ \emph {et~al.}(2018)\citenamefont {{Del
  Pino}}, \citenamefont {Schr{\"{o}}der}, \citenamefont {Chin}, \citenamefont
  {Feist},\ and\ \citenamefont {Garcia-Vidal}}]{DelPino2018}%
  \BibitemOpen
  \bibfield  {author} {\bibinfo {author} {\bibfnamefont {J.}~\bibnamefont {{Del
  Pino}}}, \bibinfo {author} {\bibfnamefont {F.~A.}\ \bibnamefont
  {Schr{\"{o}}der}}, \bibinfo {author} {\bibfnamefont {A.~W.}\ \bibnamefont
  {Chin}}, \bibinfo {author} {\bibfnamefont {J.}~\bibnamefont {Feist}},\ and\
  \bibinfo {author} {\bibfnamefont {F.~J.}\ \bibnamefont {Garcia-Vidal}},\
  }\href {https://doi.org/10.1103/PhysRevLett.121.227401} {\bibfield  {journal}
  {\bibinfo  {journal} {Physical Review Letters}\ }\textbf {\bibinfo {volume}
  {121}},\ \bibinfo {pages} {227401} (\bibinfo {year} {2018})},\ \Eprint
  {https://arxiv.org/abs/1804.04511} {arXiv:1804.04511} \BibitemShut {NoStop}%
\bibitem [{\citenamefont {Werner}\ \emph {et~al.}(2016)\citenamefont {Werner},
  \citenamefont {Jaschke}, \citenamefont {Silvi}, \citenamefont {Kliesch},
  \citenamefont {Calarco}, \citenamefont {Eisert},\ and\ \citenamefont
  {Montangero}}]{Werner2016}%
  \BibitemOpen
  \bibfield  {author} {\bibinfo {author} {\bibfnamefont {A.~H.}\ \bibnamefont
  {Werner}}, \bibinfo {author} {\bibfnamefont {D.}~\bibnamefont {Jaschke}},
  \bibinfo {author} {\bibfnamefont {P.}~\bibnamefont {Silvi}}, \bibinfo
  {author} {\bibfnamefont {M.}~\bibnamefont {Kliesch}}, \bibinfo {author}
  {\bibfnamefont {T.}~\bibnamefont {Calarco}}, \bibinfo {author} {\bibfnamefont
  {J.}~\bibnamefont {Eisert}},\ and\ \bibinfo {author} {\bibfnamefont
  {S.}~\bibnamefont {Montangero}},\ }\href
  {https://doi.org/10.1103/PhysRevLett.116.237201} {\bibfield  {journal}
  {\bibinfo  {journal} {Physical Review Letters}\ }\textbf {\bibinfo {volume}
  {116}},\ \bibinfo {pages} {237201} (\bibinfo {year} {2016})}\BibitemShut
  {NoStop}%
\bibitem [{\citenamefont {Doria}\ \emph {et~al.}(2011)\citenamefont {Doria},
  \citenamefont {Calarco},\ and\ \citenamefont {Montangero}}]{Doria2011}%
  \BibitemOpen
  \bibfield  {author} {\bibinfo {author} {\bibfnamefont {P.}~\bibnamefont
  {Doria}}, \bibinfo {author} {\bibfnamefont {T.}~\bibnamefont {Calarco}},\
  and\ \bibinfo {author} {\bibfnamefont {S.}~\bibnamefont {Montangero}},\
  }\href {https://doi.org/10.1103/PhysRevLett.106.190501} {\bibfield  {journal}
  {\bibinfo  {journal} {Physical Review Letters}\ }\textbf {\bibinfo {volume}
  {106}},\ \bibinfo {pages} {190501} (\bibinfo {year} {2011})}\BibitemShut
  {NoStop}%
\bibitem [{\citenamefont {Bukov}\ \emph {et~al.}(2018)\citenamefont {Bukov},
  \citenamefont {Day}, \citenamefont {Sels}, \citenamefont {Weinberg},
  \citenamefont {Polkovnikov},\ and\ \citenamefont {Mehta}}]{Bukov2018}%
  \BibitemOpen
  \bibfield  {author} {\bibinfo {author} {\bibfnamefont {M.}~\bibnamefont
  {Bukov}}, \bibinfo {author} {\bibfnamefont {A.~G.}\ \bibnamefont {Day}},
  \bibinfo {author} {\bibfnamefont {D.}~\bibnamefont {Sels}}, \bibinfo {author}
  {\bibfnamefont {P.}~\bibnamefont {Weinberg}}, \bibinfo {author}
  {\bibfnamefont {A.}~\bibnamefont {Polkovnikov}},\ and\ \bibinfo {author}
  {\bibfnamefont {P.}~\bibnamefont {Mehta}},\ }\href
  {https://doi.org/10.1103/PhysRevX.8.031086} {\bibfield  {journal} {\bibinfo
  {journal} {Physical Review X}\ }\textbf {\bibinfo {volume} {8}},\ \bibinfo
  {pages} {031086} (\bibinfo {year} {2018})},\ \Eprint
  {https://arxiv.org/abs/1705.00565} {arXiv:1705.00565} \BibitemShut {NoStop}%
\bibitem [{\citenamefont {Niu}\ \emph {et~al.}(2019)\citenamefont {Niu},
  \citenamefont {Boixo}, \citenamefont {Smelyanskiy},\ and\ \citenamefont
  {Neven}}]{Niu2019}%
  \BibitemOpen
  \bibfield  {author} {\bibinfo {author} {\bibfnamefont {M.~Y.}\ \bibnamefont
  {Niu}}, \bibinfo {author} {\bibfnamefont {S.}~\bibnamefont {Boixo}}, \bibinfo
  {author} {\bibfnamefont {V.~N.}\ \bibnamefont {Smelyanskiy}},\ and\ \bibinfo
  {author} {\bibfnamefont {H.}~\bibnamefont {Neven}},\ }\href
  {https://doi.org/10.1038/s41534-019-0141-3} {\bibfield  {journal} {\bibinfo
  {journal} {npj Quantum Information}\ }\textbf {\bibinfo {volume} {5}},\
  \bibinfo {pages} {1} (\bibinfo {year} {2019})}\BibitemShut {NoStop}%
\bibitem [{\citenamefont {Wang}\ \emph {et~al.}(2019)\citenamefont {Wang},
  \citenamefont {Qiu}, \citenamefont {Xiao}, \citenamefont {Zhan},
  \citenamefont {Bian}, \citenamefont {Yi},\ and\ \citenamefont
  {Xue}}]{Wang2019}%
  \BibitemOpen
  \bibfield  {author} {\bibinfo {author} {\bibfnamefont {K.}~\bibnamefont
  {Wang}}, \bibinfo {author} {\bibfnamefont {X.}~\bibnamefont {Qiu}}, \bibinfo
  {author} {\bibfnamefont {L.}~\bibnamefont {Xiao}}, \bibinfo {author}
  {\bibfnamefont {X.}~\bibnamefont {Zhan}}, \bibinfo {author} {\bibfnamefont
  {Z.}~\bibnamefont {Bian}}, \bibinfo {author} {\bibfnamefont {W.}~\bibnamefont
  {Yi}},\ and\ \bibinfo {author} {\bibfnamefont {P.}~\bibnamefont {Xue}},\
  }\href {https://doi.org/10.1103/PhysRevLett.122.020501} {\bibfield  {journal}
  {\bibinfo  {journal} {Physical Review Letters}\ }\textbf {\bibinfo {volume}
  {122}},\ \bibinfo {pages} {020501} (\bibinfo {year} {2019})},\ \Eprint
  {https://arxiv.org/abs/1806.10871} {arXiv:1806.10871} \BibitemShut {NoStop}%
\bibitem [{\citenamefont {Worth}\ \emph {et~al.}(2008)\citenamefont {Worth},
  \citenamefont {Meyer}, \citenamefont {K{\"{o}}ppel}, \citenamefont
  {Cederbaum},\ and\ \citenamefont {Burghardt}}]{Worth2008}%
  \BibitemOpen
  \bibfield  {author} {\bibinfo {author} {\bibfnamefont {G.~A.}\ \bibnamefont
  {Worth}}, \bibinfo {author} {\bibfnamefont {H.-D.}\ \bibnamefont {Meyer}},
  \bibinfo {author} {\bibfnamefont {H.}~\bibnamefont {K{\"{o}}ppel}}, \bibinfo
  {author} {\bibfnamefont {L.~S.}\ \bibnamefont {Cederbaum}},\ and\ \bibinfo
  {author} {\bibfnamefont {I.}~\bibnamefont {Burghardt}},\ }\href
  {https://doi.org/10.1080/01442350802137656} {\bibfield  {journal} {\bibinfo
  {journal} {International Reviews in Physical Chemistry}\ }\textbf {\bibinfo
  {volume} {27}},\ \bibinfo {pages} {569} (\bibinfo {year} {2008})}\BibitemShut
  {NoStop}%
\bibitem [{\citenamefont {Schollw{\"{o}}ck}(2011)}]{Schollwock2011}%
  \BibitemOpen
  \bibfield  {author} {\bibinfo {author} {\bibfnamefont {U.}~\bibnamefont
  {Schollw{\"{o}}ck}},\ }\href {https://doi.org/10.1016/j.aop.2010.09.012}
  {\bibfield  {journal} {\bibinfo  {journal} {Annals of Physics}\ }\textbf
  {\bibinfo {volume} {326}},\ \bibinfo {pages} {96} (\bibinfo {year} {2011})},\
  \Eprint {https://arxiv.org/abs/1008.3477} {arXiv:1008.3477} \BibitemShut
  {NoStop}%
\bibitem [{\citenamefont {Verstraete}\ \emph {et~al.}(2008)\citenamefont
  {Verstraete}, \citenamefont {Murg},\ and\ \citenamefont
  {Cirac}}]{Verstraete2008}%
  \BibitemOpen
  \bibfield  {author} {\bibinfo {author} {\bibfnamefont {F.}~\bibnamefont
  {Verstraete}}, \bibinfo {author} {\bibfnamefont {V.}~\bibnamefont {Murg}},\
  and\ \bibinfo {author} {\bibfnamefont {J.}~\bibnamefont {Cirac}},\ }\href
  {https://doi.org/10.1080/14789940801912366} {\bibfield  {journal} {\bibinfo
  {journal} {Advances in Physics}\ }\textbf {\bibinfo {volume} {57}},\ \bibinfo
  {pages} {143} (\bibinfo {year} {2008})}\BibitemShut {NoStop}%
\bibitem [{\citenamefont {Vidal}(2007)}]{Vidal2007}%
  \BibitemOpen
  \bibfield  {author} {\bibinfo {author} {\bibfnamefont {G.}~\bibnamefont
  {Vidal}},\ }\href {https://doi.org/10.1103/PhysRevLett.99.220405} {\bibfield
  {journal} {\bibinfo  {journal} {Physical Review Letters}\ }\textbf {\bibinfo
  {volume} {99}},\ \bibinfo {pages} {220405} (\bibinfo {year} {2007})},\
  \Eprint {https://arxiv.org/abs/0512165} {arXiv:0512165 [cond-mat]}
  \BibitemShut {NoStop}%
\bibitem [{\citenamefont {White}(1992)}]{White1992}%
  \BibitemOpen
  \bibfield  {author} {\bibinfo {author} {\bibfnamefont {S.~R.}\ \bibnamefont
  {White}},\ }\href {https://doi.org/10.1103/PhysRevLett.69.2863} {\bibfield
  {journal} {\bibinfo  {journal} {Physical Review Letters}\ }\textbf {\bibinfo
  {volume} {69}},\ \bibinfo {pages} {2863} (\bibinfo {year}
  {1992})}\BibitemShut {NoStop}%
\bibitem [{\citenamefont {Hastings}(2007)}]{Hastings2007}%
  \BibitemOpen
  \bibfield  {author} {\bibinfo {author} {\bibfnamefont {M.~B.}\ \bibnamefont
  {Hastings}},\ }\href
  {https://iopscience.iop.org/article/10.1088/1742-5468/2007/08/P08024
  https://iopscience.iop.org/article/10.1088/1742-5468/2007/08/P08024/meta}
  {\bibfield  {journal} {\bibinfo  {journal} {Journal of Statistical Mechanics:
  Theory and Experiment}\ }\textbf {\bibinfo {volume} {2007}},\ \bibinfo
  {pages} {P08024} (\bibinfo {year} {2007})}\BibitemShut {NoStop}%
\bibitem [{\citenamefont {Bravyi}\ \emph {et~al.}(2006)\citenamefont {Bravyi},
  \citenamefont {Hastings},\ and\ \citenamefont {Verstraete}}]{Bravyi2006}%
  \BibitemOpen
  \bibfield  {author} {\bibinfo {author} {\bibfnamefont {S.}~\bibnamefont
  {Bravyi}}, \bibinfo {author} {\bibfnamefont {M.~B.}\ \bibnamefont
  {Hastings}},\ and\ \bibinfo {author} {\bibfnamefont {F.}~\bibnamefont
  {Verstraete}},\ }\href {https://doi.org/10.1103/PhysRevLett.97.050401}
  {\bibfield  {journal} {\bibinfo  {journal} {Physical Review Letters}\
  }\textbf {\bibinfo {volume} {97}},\ \bibinfo {pages} {050401} (\bibinfo
  {year} {2006})},\ \Eprint {https://arxiv.org/abs/0603121} {arXiv:0603121
  [quant-ph]} \BibitemShut {NoStop}%
\bibitem [{\citenamefont {Deng}\ \emph {et~al.}(2017)\citenamefont {Deng},
  \citenamefont {Li},\ and\ \citenamefont {{Das Sarma}}}]{Deng2017}%
  \BibitemOpen
  \bibfield  {author} {\bibinfo {author} {\bibfnamefont {D.~L.}\ \bibnamefont
  {Deng}}, \bibinfo {author} {\bibfnamefont {X.}~\bibnamefont {Li}},\ and\
  \bibinfo {author} {\bibfnamefont {S.}~\bibnamefont {{Das Sarma}}},\ }\href
  {https://doi.org/10.1103/PhysRevX.7.021021} {\bibinfo {title} {{Quantum
  entanglement in neural network states}}} (\bibinfo {year} {2017}),\ \Eprint
  {https://arxiv.org/abs/1701.04844} {arXiv:1701.04844} \BibitemShut {NoStop}%
\bibitem [{\citenamefont {Carleo}\ and\ \citenamefont
  {Troyer}(2017)}]{Carleo2017}%
  \BibitemOpen
  \bibfield  {author} {\bibinfo {author} {\bibfnamefont {G.}~\bibnamefont
  {Carleo}}\ and\ \bibinfo {author} {\bibfnamefont {M.}~\bibnamefont
  {Troyer}},\ }\href {https://doi.org/10.1126/science.aag2302} {\bibfield
  {journal} {\bibinfo  {journal} {Science}\ }\textbf {\bibinfo {volume}
  {355}},\ \bibinfo {pages} {602} (\bibinfo {year} {2017})}\BibitemShut
  {NoStop}%
\bibitem [{\citenamefont {Han}\ \emph {et~al.}(2019)\citenamefont {Han},
  \citenamefont {Zhang},\ and\ \citenamefont {Weinan}}]{han2019solving}%
  \BibitemOpen
  \bibfield  {author} {\bibinfo {author} {\bibfnamefont {J.}~\bibnamefont
  {Han}}, \bibinfo {author} {\bibfnamefont {L.}~\bibnamefont {Zhang}},\ and\
  \bibinfo {author} {\bibfnamefont {E.}~\bibnamefont {Weinan}},\ }\href@noop {}
  {\bibfield  {journal} {\bibinfo  {journal} {Journal of Computational
  Physics}\ }\textbf {\bibinfo {volume} {399}},\ \bibinfo {pages} {108929}
  (\bibinfo {year} {2019})}\BibitemShut {NoStop}%
\bibitem [{\citenamefont {Choo}\ \emph {et~al.}(2019)\citenamefont {Choo},
  \citenamefont {Neupert},\ and\ \citenamefont {Carleo}}]{Choo2019}%
  \BibitemOpen
  \bibfield  {author} {\bibinfo {author} {\bibfnamefont {K.}~\bibnamefont
  {Choo}}, \bibinfo {author} {\bibfnamefont {T.}~\bibnamefont {Neupert}},\ and\
  \bibinfo {author} {\bibfnamefont {G.}~\bibnamefont {Carleo}},\ }\href
  {https://doi.org/10.1103/PhysRevB.100.125124} {\bibfield  {journal} {\bibinfo
   {journal} {Physical Review B}\ }\textbf {\bibinfo {volume} {100}},\ \bibinfo
  {pages} {125124} (\bibinfo {year} {2019})}\BibitemShut {NoStop}%
\bibitem [{\citenamefont {Pfau}\ \emph {et~al.}(2019)\citenamefont {Pfau},
  \citenamefont {Spencer}, \citenamefont {Matthews},\ and\ \citenamefont
  {Foulkes}}]{Pfau2019}%
  \BibitemOpen
  \bibfield  {author} {\bibinfo {author} {\bibfnamefont {D.}~\bibnamefont
  {Pfau}}, \bibinfo {author} {\bibfnamefont {J.~S.}\ \bibnamefont {Spencer}},
  \bibinfo {author} {\bibfnamefont {A.~G. d.~G.}\ \bibnamefont {Matthews}},\
  and\ \bibinfo {author} {\bibfnamefont {W.~M.~C.}\ \bibnamefont {Foulkes}},\
  }\href@noop {} {\bibfield  {journal} {\bibinfo  {journal} {arXiv preprint
  arXiv:1909.02487}\ } (\bibinfo {year} {2019})}\BibitemShut {NoStop}%
\bibitem [{\citenamefont {Luo}\ and\ \citenamefont
  {Clark}(2019)}]{luo2019backflow}%
  \BibitemOpen
  \bibfield  {author} {\bibinfo {author} {\bibfnamefont {D.}~\bibnamefont
  {Luo}}\ and\ \bibinfo {author} {\bibfnamefont {B.~K.}\ \bibnamefont
  {Clark}},\ }\href@noop {} {\bibfield  {journal} {\bibinfo  {journal}
  {Physical review letters}\ }\textbf {\bibinfo {volume} {122}},\ \bibinfo
  {pages} {226401} (\bibinfo {year} {2019})}\BibitemShut {NoStop}%
\bibitem [{\citenamefont {Hermann}\ \emph {et~al.}(2019)\citenamefont
  {Hermann}, \citenamefont {Sch{\"a}tzle},\ and\ \citenamefont
  {No{\'e}}}]{hermann2019deep}%
  \BibitemOpen
  \bibfield  {author} {\bibinfo {author} {\bibfnamefont {J.}~\bibnamefont
  {Hermann}}, \bibinfo {author} {\bibfnamefont {Z.}~\bibnamefont
  {Sch{\"a}tzle}},\ and\ \bibinfo {author} {\bibfnamefont {F.}~\bibnamefont
  {No{\'e}}},\ }\href@noop {} {\bibfield  {journal} {\bibinfo  {journal} {arXiv
  preprint arXiv:1909.08423}\ } (\bibinfo {year} {2019})}\BibitemShut {NoStop}%
\bibitem [{\citenamefont {Levine}\ \emph {et~al.}(2019)\citenamefont {Levine},
  \citenamefont {Sharir}, \citenamefont {Cohen},\ and\ \citenamefont
  {Shashua}}]{Levine2019}%
  \BibitemOpen
  \bibfield  {author} {\bibinfo {author} {\bibfnamefont {Y.}~\bibnamefont
  {Levine}}, \bibinfo {author} {\bibfnamefont {O.}~\bibnamefont {Sharir}},
  \bibinfo {author} {\bibfnamefont {N.}~\bibnamefont {Cohen}},\ and\ \bibinfo
  {author} {\bibfnamefont {A.}~\bibnamefont {Shashua}},\ }\href
  {https://doi.org/10.1103/PhysRevLett.122.065301} {\bibfield  {journal}
  {\bibinfo  {journal} {Physical Review Letters}\ }\textbf {\bibinfo {volume}
  {122}},\ \bibinfo {pages} {065301} (\bibinfo {year} {2019})},\ \Eprint
  {https://arxiv.org/abs/1803.09780} {arXiv:1803.09780} \BibitemShut {NoStop}%
\bibitem [{\citenamefont {Schmitt}\ and\ \citenamefont {Heyl}(2019)}]{Schmitt}%
  \BibitemOpen
  \bibfield  {author} {\bibinfo {author} {\bibfnamefont {M.}~\bibnamefont
  {Schmitt}}\ and\ \bibinfo {author} {\bibfnamefont {M.}~\bibnamefont {Heyl}},\
  }\href@noop {} {\bibfield  {journal} {\bibinfo  {journal} {arXiv preprint
  arXiv:1912.08828}\ } (\bibinfo {year} {2019})}\BibitemShut {NoStop}%
\bibitem [{\citenamefont {Fabiani}\ and\ \citenamefont
  {Mentink}(2019)}]{Fabiani}%
  \BibitemOpen
  \bibfield  {author} {\bibinfo {author} {\bibfnamefont {G.}~\bibnamefont
  {Fabiani}}\ and\ \bibinfo {author} {\bibfnamefont {J.}~\bibnamefont
  {Mentink}},\ }\href@noop {} {\bibfield  {journal} {\bibinfo  {journal} {arXiv
  preprint arXiv:1912.10845}\ } (\bibinfo {year} {2019})}\BibitemShut {NoStop}%
\bibitem [{\citenamefont {Hartmann}\ and\ \citenamefont
  {Carleo}(2019)}]{Hartmann2019}%
  \BibitemOpen
  \bibfield  {author} {\bibinfo {author} {\bibfnamefont {M.~J.}\ \bibnamefont
  {Hartmann}}\ and\ \bibinfo {author} {\bibfnamefont {G.}~\bibnamefont
  {Carleo}},\ }\href@noop {} {\bibfield  {journal} {\bibinfo  {journal}
  {Physical review letters}\ }\textbf {\bibinfo {volume} {122}},\ \bibinfo
  {pages} {250502} (\bibinfo {year} {2019})}\BibitemShut {NoStop}%
\bibitem [{\citenamefont {Yoshioka}\ and\ \citenamefont
  {Hamazaki}(2019)}]{yoshioka2019constructing}%
  \BibitemOpen
  \bibfield  {author} {\bibinfo {author} {\bibfnamefont {N.}~\bibnamefont
  {Yoshioka}}\ and\ \bibinfo {author} {\bibfnamefont {R.}~\bibnamefont
  {Hamazaki}},\ }\href@noop {} {\bibfield  {journal} {\bibinfo  {journal}
  {Physical Review B}\ }\textbf {\bibinfo {volume} {99}},\ \bibinfo {pages}
  {214306} (\bibinfo {year} {2019})}\BibitemShut {NoStop}%
\bibitem [{\citenamefont {Kosut}\ \emph {et~al.}(2020)\citenamefont {Kosut},
  \citenamefont {Ho},\ and\ \citenamefont {Rabitz}}]{Kosut2020}%
  \BibitemOpen
  \bibfield  {author} {\bibinfo {author} {\bibfnamefont {R.~L.}\ \bibnamefont
  {Kosut}}, \bibinfo {author} {\bibfnamefont {T.-S.}\ \bibnamefont {Ho}},\ and\
  \bibinfo {author} {\bibfnamefont {H.}~\bibnamefont {Rabitz}},\ }\href@noop {}
  {\bibfield  {journal} {\bibinfo  {journal} {arXiv preprint arXiv:2006.13498}\
  } (\bibinfo {year} {2020})}\BibitemShut {NoStop}%
\bibitem [{\citenamefont {Alet}\ and\ \citenamefont
  {Laflorencie}(2018)}]{Alet2018}%
  \BibitemOpen
  \bibfield  {author} {\bibinfo {author} {\bibfnamefont {F.}~\bibnamefont
  {Alet}}\ and\ \bibinfo {author} {\bibfnamefont {N.}~\bibnamefont
  {Laflorencie}},\ }\href {https://doi.org/10.1016/j.crhy.2018.03.003}
  {\bibinfo {title} {{Many-body localization: An introduction and selected
  topics}}} (\bibinfo {year} {2018}),\ \Eprint
  {https://arxiv.org/abs/1711.03145} {arXiv:1711.03145} \BibitemShut {NoStop}%
\bibitem [{\citenamefont {Sharir}\ \emph {et~al.}(2020)\citenamefont {Sharir},
  \citenamefont {Levine}, \citenamefont {Wies}, \citenamefont {Carleo},\ and\
  \citenamefont {Shashua}}]{Sharir2020}%
  \BibitemOpen
  \bibfield  {author} {\bibinfo {author} {\bibfnamefont {O.}~\bibnamefont
  {Sharir}}, \bibinfo {author} {\bibfnamefont {Y.}~\bibnamefont {Levine}},
  \bibinfo {author} {\bibfnamefont {N.}~\bibnamefont {Wies}}, \bibinfo {author}
  {\bibfnamefont {G.}~\bibnamefont {Carleo}},\ and\ \bibinfo {author}
  {\bibfnamefont {A.}~\bibnamefont {Shashua}},\ }\href
  {https://doi.org/10.1103/PhysRevLett.124.020503} {\bibfield  {journal}
  {\bibinfo  {journal} {Physical Review Letters}\ }\textbf {\bibinfo {volume}
  {124}},\ \bibinfo {pages} {020503} (\bibinfo {year} {2020})},\ \Eprint
  {https://arxiv.org/abs/1902.04057} {arXiv:1902.04057} \BibitemShut {NoStop}%
\bibitem [{\citenamefont {Wahl}\ \emph
  {et~al.}(2017{\natexlab{b}})\citenamefont {Wahl}, \citenamefont {Pal},\ and\
  \citenamefont {Simon}}]{wahl2017efficient}%
  \BibitemOpen
  \bibfield  {author} {\bibinfo {author} {\bibfnamefont {T.~B.}\ \bibnamefont
  {Wahl}}, \bibinfo {author} {\bibfnamefont {A.}~\bibnamefont {Pal}},\ and\
  \bibinfo {author} {\bibfnamefont {S.~H.}\ \bibnamefont {Simon}},\ }\href@noop
  {} {\bibfield  {journal} {\bibinfo  {journal} {Physical Review X}\ }\textbf
  {\bibinfo {volume} {7}},\ \bibinfo {pages} {021018} (\bibinfo {year}
  {2017}{\natexlab{b}})}\BibitemShut {NoStop}%
\bibitem [{\citenamefont {Czischek}\ \emph {et~al.}(2018)\citenamefont
  {Czischek}, \citenamefont {G{\"{a}}rttner},\ and\ \citenamefont
  {Gasenzer}}]{Czischek2018}%
  \BibitemOpen
  \bibfield  {author} {\bibinfo {author} {\bibfnamefont {S.}~\bibnamefont
  {Czischek}}, \bibinfo {author} {\bibfnamefont {M.}~\bibnamefont
  {G{\"{a}}rttner}},\ and\ \bibinfo {author} {\bibfnamefont {T.}~\bibnamefont
  {Gasenzer}},\ }\href {https://doi.org/10.1103/PhysRevB.98.024311} {\bibfield
  {journal} {\bibinfo  {journal} {Physical Review B}\ }\textbf {\bibinfo
  {volume} {98}},\ \bibinfo {pages} {024311} (\bibinfo {year} {2018})},\
  \Eprint {https://arxiv.org/abs/1803.08321} {arXiv:1803.08321} \BibitemShut
  {NoStop}%
\bibitem [{\citenamefont {Oord}\ \emph {et~al.}(2016)\citenamefont {Oord},
  \citenamefont {Dieleman}, \citenamefont {Zen}, \citenamefont {Simonyan},
  \citenamefont {Vinyals}, \citenamefont {Graves}, \citenamefont
  {Kalchbrenner}, \citenamefont {Senior},\ and\ \citenamefont
  {Kavukcuoglu}}]{Oord2016}%
  \BibitemOpen
  \bibfield  {author} {\bibinfo {author} {\bibfnamefont {A.~v.~d.}\
  \bibnamefont {Oord}}, \bibinfo {author} {\bibfnamefont {S.}~\bibnamefont
  {Dieleman}}, \bibinfo {author} {\bibfnamefont {H.}~\bibnamefont {Zen}},
  \bibinfo {author} {\bibfnamefont {K.}~\bibnamefont {Simonyan}}, \bibinfo
  {author} {\bibfnamefont {O.}~\bibnamefont {Vinyals}}, \bibinfo {author}
  {\bibfnamefont {A.}~\bibnamefont {Graves}}, \bibinfo {author} {\bibfnamefont
  {N.}~\bibnamefont {Kalchbrenner}}, \bibinfo {author} {\bibfnamefont
  {A.}~\bibnamefont {Senior}},\ and\ \bibinfo {author} {\bibfnamefont
  {K.}~\bibnamefont {Kavukcuoglu}},\ }\href@noop {} {\bibfield  {journal}
  {\bibinfo  {journal} {arXiv preprint arXiv:1609.03499}\ } (\bibinfo {year}
  {2016})}\BibitemShut {NoStop}%
\bibitem [{\citenamefont {Paszke}\ \emph {et~al.}(2019)\citenamefont {Paszke},
  \citenamefont {Gross}, \citenamefont {Massa}, \citenamefont {Lerer},
  \citenamefont {Bradbury}, \citenamefont {Chanan}, \citenamefont {Killeen},
  \citenamefont {Lin}, \citenamefont {Gimelshein}, \citenamefont {Antiga} \emph
  {et~al.}}]{Paszke2019}%
  \BibitemOpen
  \bibfield  {author} {\bibinfo {author} {\bibfnamefont {A.}~\bibnamefont
  {Paszke}}, \bibinfo {author} {\bibfnamefont {S.}~\bibnamefont {Gross}},
  \bibinfo {author} {\bibfnamefont {F.}~\bibnamefont {Massa}}, \bibinfo
  {author} {\bibfnamefont {A.}~\bibnamefont {Lerer}}, \bibinfo {author}
  {\bibfnamefont {J.}~\bibnamefont {Bradbury}}, \bibinfo {author}
  {\bibfnamefont {G.}~\bibnamefont {Chanan}}, \bibinfo {author} {\bibfnamefont
  {T.}~\bibnamefont {Killeen}}, \bibinfo {author} {\bibfnamefont
  {Z.}~\bibnamefont {Lin}}, \bibinfo {author} {\bibfnamefont {N.}~\bibnamefont
  {Gimelshein}}, \bibinfo {author} {\bibfnamefont {L.}~\bibnamefont {Antiga}},
  \emph {et~al.},\ }in\ \href@noop {} {\emph {\bibinfo {booktitle} {Advances in
  neural information processing systems}}}\ (\bibinfo {year} {2019})\ pp.\
  \bibinfo {pages} {8026--8037}\BibitemShut {NoStop}%
\bibitem [{\citenamefont {Kingma}\ and\ \citenamefont
  {Ba}(2014)}]{kingma2014adam}%
  \BibitemOpen
  \bibfield  {author} {\bibinfo {author} {\bibfnamefont {D.~P.}\ \bibnamefont
  {Kingma}}\ and\ \bibinfo {author} {\bibfnamefont {J.}~\bibnamefont {Ba}},\
  }\href@noop {} {\bibfield  {journal} {\bibinfo  {journal} {arXiv preprint
  arXiv:1412.6980}\ } (\bibinfo {year} {2014})}\BibitemShut {NoStop}%
\end{thebibliography}%
\end{document}